\documentclass[twoside,11pt]{article}
\usepackage{jmlr2e}

\ShortHeadings{FedNorm: Modality-Based Normalization in Federated Learning}{Bernecker et al.}
\firstpageno{1}

\usepackage{amsmath,amssymb,amsfonts}
\usepackage[noend]{algorithmic}
\usepackage{graphicx}
\usepackage{textcomp}

\usepackage{booktabs}
\usepackage[counterclockwise]{rotating}
\usepackage{arydshln}
\usepackage{adjustbox}
\usepackage{algorithm}
\usepackage{subfig}
\usepackage{cleveref}
\usepackage{xspace}
\usepackage{hyphenat}
\usepackage{url}
\usepackage{multirow}

\newcommand{\fedavg}{\texttt{FedAvg}\xspace}
\newcommand{\fedavgm}{\texttt{FedAvgM}\xspace}
\newcommand{\fedvc}{\texttt{FedVC}\xspace}
\newcommand{\silobn}{\texttt{SiloBN}\xspace}
\newcommand{\fedbn}{\texttt{FedBN}\xspace}
\newcommand{\fednorm}{\texttt{FedNorm}\xspace}
\newcommand{\fednormp}{\texttt{FedNorm+}\xspace}

\newcommand{\SUB}[1]{\STATE \textbf{#1}}

\begin{document}

\title{FedNorm: Modality-Based Normalization in Federated Learning for Multi-Modal Liver Segmentation}

\author{\name Tobias Bernecker$^{1,2}$,
\name Annette Peters$^{1}$,
\name Christopher L. Schlett$^{3}$,
\name Fabian Bamberg$^{3}$,
\name Fabian Theis$^{1}$,
\name Daniel Rueckert$^{2}$,
\name Jakob Weiß$^{3}$ and
\name Shadi Albarqouni$^{1,2,4,\dagger}$\\
\addr $^{1}$ Helmholtz Zentrum München, Neuherberg, Germany \\
\addr $^{2}$ Technical University of Munich, Garching, Germany\\
\addr $^{3}$ Medical Center-University of Freiburg, University of Freiburg, Freiburg, Germany\\
\addr $^{4}$ University Hospital Bonn, University of Bonn, Bonn, Germany\\ 
\addr $^{\dagger}$ Corresponding Author: (e-mail: shadi.albarqouni@ukbonn.de)}


\editor{}

\maketitle

\begin{abstract}
Given the high incidence and effective treatment options for liver diseases, they are of great socioeconomic importance. One of the most common methods for analyzing CT and MRI images for diagnosis and follow-up treatment is liver segmentation. Recent advances in deep learning have demonstrated encouraging results for automatic liver segmentation. Despite this, their success depends primarily on the availability of an annotated database, which is often not available because of privacy concerns. Federated Learning has been recently proposed as a solution to alleviate these challenges by training a shared global model on distributed clients without access to their local databases. Nevertheless, Federated Learning does not perform well when it is trained on a high degree of heterogeneity of image data due to multi-modal imaging, such as CT and MRI, and multiple scanner types. To this end, we propose \fednorm and its extension \fednormp, two Federated Learning algorithms that use a modality-based normalization technique. Specifically,  \fednorm normalizes the features on a client-level, while \fednormp employs the modality information of single slices in the feature normalization. Our methods were validated using 428 patients from six publicly available databases and compared to state-of-the-art Federated Learning algorithms and baseline models in heterogeneous settings (multi-institutional, multi-modal data). The experimental results demonstrate that our methods show an overall acceptable performance, achieve Dice per patient scores up to 0.961, consistently outperform locally trained models, and are on par or slightly better than centralized models.
\end{abstract}

\section{Introduction}\label{sec:introduction}

Liver cancer is the second most common cause of cancer deaths in the world with 830,180 in 
2020 according to the Global Cancer Observatory~\citep{LiverCancer}.
Liver diseases such as hepatic steatosis, liver fibrosis as well as primary and secondary malignancies are of great socioeconomic importance given their high incidence and effective treatment options~\citep{LiverDiseases}. Thus, early diagnosis is desirable to reduce morbidity and mortality. Recent advanced in artificial intelligence, deep learning in particular, have shown encouraging results for automatic detection and quantification of these conditions.
Such models are often trained in a supervised fashion, which requires a large amount of pixel-wise 
annotated data~\citep{UNet, TrainingDataSize}. Collecting ground-truth annotation is oftentimes time-consuming, tedious, and labor 
expensive. Besides, such a large database might not be available in a single institute or hospital due to 
the low incidence rate of pathological cases~\citep{PrivacyPreservingFL}. 

A straightforward approach to overcome the aforementioned  challenges would be to train a model on a collection of available training data (data lake) from 
multiple institutions. However, sharing sensitive data is often not possible, especially in the medical 
sector, due to data protection laws~\citep{PrivacyPreservingFL} such as 
the General Data Protection 
Regulation~\footnote{\url{http://data.europa.eu/eli/reg/2016/679/oj}. 
(Accessed: 2021-05-29)}
Therefore, it would be helpful to train a shared global model in a 
decentralized fashion
based on the available training data from multiple hospitals and institutes.

In 2017, McMahan~et~al.~\citep{FedAvg} introduced the concept of Federated Learning (FL) where they 
demonstrated that training of a shared global model can be performed in a decentralized way, which 
overcomes the problem of data sharing. Since then, multiple works focused on heterogeneity in
FL~\citep{SiloBN, FederatedMMB, FedVC, FedAvgM, FedProx, FedBN, FLNonIID}.
FL provides an adequate solution for training deep learning models with data coming from multiple 
institutions, especially in privacy-sensitive domains like medical
imaging~\citep{PrivacyPreservingFLMI, PrivacyPreservingFLMI2, PrivacyPreservingFL, PrivacyPreservingFL2, MedicalFL}.
In this context, additional focus was placed on privacy-preserving
FL~\citep{PrivacyPreservingFLMI, PrivacyPreservingFLMI2, PrivacyPreservingFL, PrivacyPreservingFL2}.
Nevertheless, FL was already used to train models for medical image segmentation, in many cases for brain
tumor segmentation~\citep{PrivacyPreservingFL, MedicalFL}.

However, training a shared global model for medical image segmentation in a federated fashion poses several challenges. 
These include training deep learning models on multiple imaging modalities with a large number of scanners from 
different manufacturers, various image resolutions and protocols. This data can be seen as not independent and 
identically distributed (non-IID)~\citep{DataHeterogeneity}. Furthermore, there is an imbalance of abundant 
annotated computed tomography (CT) scans and rarely annotated magnetic resonance imaging (MRI) scans, due to differences in acquisition times and
costs~\citep{LiverCancerImagingModalities, DADR}. For this reason, the CT and MRI segmentation performance could 
benefit from training a model on both imaging modalities~\citep{MMOS}. In the past, multiple works focused on 
training liver/organ segmentation models on scans from multi-modal (CT and MRI) data~\citep{MMOS, MMLSGCNN, DADR}, 
on different MRI sequences~\citep{MMLSMRI} or for multi-modal liver lesion
segmentation~\citep{MultiModalLiverLesionSegmentation}. Other works propose methods that are trained on one modality
and can generalize to the other modality using domain
adaptation~\citep{MultiModalMedicalImageSegmentation, CrossModalityUnsupervisedOrganSegmentation, DADR}. However, 
many works only focus on liver/organ segmentation on a single
modality~\citep{LiverSegmentationGraphCut2, CascadedFCN, nnUNet, HDenseUNet, LiverSegmentationGraphCut, VNet},
or propose methods that perform well for both imaging modalities, but are trained and tested separately for each 
modality~\citep{MMLTSCascadedFCN, MMLS}.

In this work, we aim to answer the question whether deep learning models can be trained on CT and MRI data for 
multi-modal liver segmentation in a federated fashion. To this end, we present two novel FL approaches, \fednorm and 
\fednormp, which are based on the Mode Normalization (MN)~\citep{ModeNormalization} technique and inspired by the 
Federated Averaging with Server Momentum (\fedavgm)~\citep{FedAvgM} algorithm. To the best of our knowledge, this 
work is the first work to train liver segmentation models on multi-modal data with FL. In our experiments, we compare 
\fednorm and \fednormp to baselines and other state-of-the-art FL algorithms.

Even if privacy preservation plays a key role in FL with medical
data~\citep{PrivacyPreservingFLMI, PrivacyPreservingFLMI2, PrivacyPreservingFL, PrivacyPreservingFL2}, this work 
rather focuses on overcoming the problem of training deep learning models for multi-modal liver segmentation in a 
decentralized fashion, by reaching similar or better performance than models trained on data lakes.

The rest of this paper has the following structure. \Cref{sec:related-work} gives an overview about different 
approaches for multi-modal liver segmentation and FL algorithms, which are designed to tackle the problem of data 
heterogeneity~\citep{DataHeterogeneity} in FL. \Cref{sec:methodology} introduces our new approach, \fednorm, and
its extension, \fednormp, for multi-modal liver segmentation with FL. In \Cref{sec:experiments-results}, the 
datasets used in this work are presented. Furthermore, our experiments are explained in more detail and the corresponding results are shown. Finally, conclusions are drawn in \Cref{sec:conclusion}.
\section{Related Work}\label{sec:related-work}
This work focuses on multi-modal liver segmentation on CT and MRI data with FL. Hence, this section 
highlights the research areas of multi-modal liver/organ segmentation on CT and MRI data, and FL with 
data heterogeneity~\citep{DataHeterogeneity} by presenting relevant methods in more detail.

\paragraph{Multi-Modal Liver Segmentation}
Training deep learning models with scans from multiple modalities, e.g. CT and MRI, can be challenging 
due to the domain shift between the modalities caused by differences in the scanners and protocols, 
which are used to acquire the medical images (see \Cref{sec:challenges})~\citep{DADR}.
On the other hand, training a model on scans from multiple modalities increases the amount of available
training data, which may not be available in sufficient quantity for a specific modality, and leads to 
a better segmentation performance on the single modalities~\citep{MMOS}. In this sense an early work used an 
encoder-decoder network to extract shared information from unpaired CT and MRI scans to perform multi-modal organ segmentation~\citep{MMOS}.
Another method uses transfer learning to refine an initial model, which is pre-trained on a single modality, for CT and MRI liver segmentation~\citep{MMLSGCNN}.
Usually, domain adaptation can be used to overcome the problem of domain shift between CT and MRI data.
To this end, one approach proposes to learn disentangled representations in an unsupervised way to 
separate the content and style representations of CT and MRI scans~\citep{DADR}. Afterwards, a 
U-Net~\citep{UNet} is trained for the task of liver segmentation on generated content-only images of the
respective CT and MRI scans~\citep{DADR}.
However, all of these approaches train the networks in a centralized fashion and require both 
modalities during training. Therefore, they are not applicable to real-world FL settings, where both 
modalities may not be available at each client. Nevertheless, the pre-trained model from the transfer 
learning approach of Wang~et~al.~\citep{MMLSGCNN} can be refined on single modalities.
In this work, we train multi-modal liver segmentation models with FL, where each client only needs to have data from one modality, but may have data from both modalities. Thus, our methods can be applied to scenarios with real-world hospitals.

\paragraph{Federated Learning with Data Heterogeneity}
\fedavg~\citep{FedAvg}, the base algorithm for FL, can achieve a good performance on heterogeneous data 
settings~\citep{FedAvg}. It may, however, show a performance degradation on non-IID 
data~\citep{FLNonIID}. Nevertheless, \fedavg can converge to the optimum under certain circumstances 
when applied to these tasks~\citep{FedAvgNonIID}. To avoid this performance degradation and to prevent 
parameters from oscillating, the \fedavgm~algorithm~\citep{FedAvgM} uses momentum-based updates in the aggregation step of the server~\citep{FedAvgM}. However, it is very sensitive to the choice of the momentum parameter and learning 
rate~\citep{FedAvgM}.
In contrast, we propose to use an interpolation between the old state and new averaged state only of 
specific parameters, e.g. only of the normalization parameters in the network (see \Cref{sec:fednorm}),
as these parameters are the critical part. If the normalization works as intended, the other parameters should get refined to the same optimum by all clients.

Federated Virtual Clients (\fedvc)~\citep{FedVC} is an algorithm that solves the problem of different sizes of training datasets at 
each client. Therefore, it limits the number of local client updates to be the same for each client and
selects clients with a probability based on their amount of training samples~\citep{FedVC}. One drawback
of \fedvc is that clients with rare samples of data, but a low training dataset size are less likely to
get selected and therefore may be neglected during FL. Our approaches, however, are based on the 
example of \fedavg, where each client performs the full number of local updates and gets selected with 
equal probability.

Recent works propose to use Batch Normalization (BN)~\citep{BatchNormalization} to address the problem 
of data heterogeneity in FL: The \silobn~\citep{SiloBN} algorithm keeps the BN statistics
($\mu$ and $\sigma$) local at each client to learn a specific model for each client. 
\fedbn~\citep{FedBN}, on the other hand, keeps the BN parameters ($\gamma$ and $\beta$) and statistics
($\mu$ and $\sigma$) local at each client, to tackle the problem of feature distribution shift among data 
from different clients~\citep{FedBN}. Due to the missing BN statistics in the shared global models of 
\silobn and \fedbn, generalization to unseen data/clients requires to compute these statistics 
first~\citep{SiloBN, FedBN}. Additionally, the shared global model of \fedbn does not contain the BN 
parameters~\citep{FedBN}. For this reason, \fedbn requires access to these parameters from other/all 
clients inside the federation when it gets tested on unseen data~\citep{FedBN}.
We, however, identify latent modes within each modality of the data. Thus, we use the MN technique~\citep{ModeNormalization}, an extension of BN that normalizes the data 
according to multiple modes~\citep{ModeNormalization}, in our approaches. Our global model shares all 
network parameters and thus does not share the limitations of \silobn and \fedbn.
\section{Methodology}\label{sec:methodology}
In this section, we present our proposed approaches for multi-modal liver 
segmentation on CT and MRI data with FL. Therefore, we first identify some
challenges and define important design requirements that we consider in 
our approaches. Finally, we introduce our approaches, \fednorm and its 
extension \fednormp.

\subsection{Challenges and Design Requirements}\label{sec:challenges}
\begin{figure}[t]
	\centering
	\begin{minipage}{\columnwidth}
	\begin{center}
	\subfloat[CT scan.]{{\includegraphics[width=0.24\textwidth]{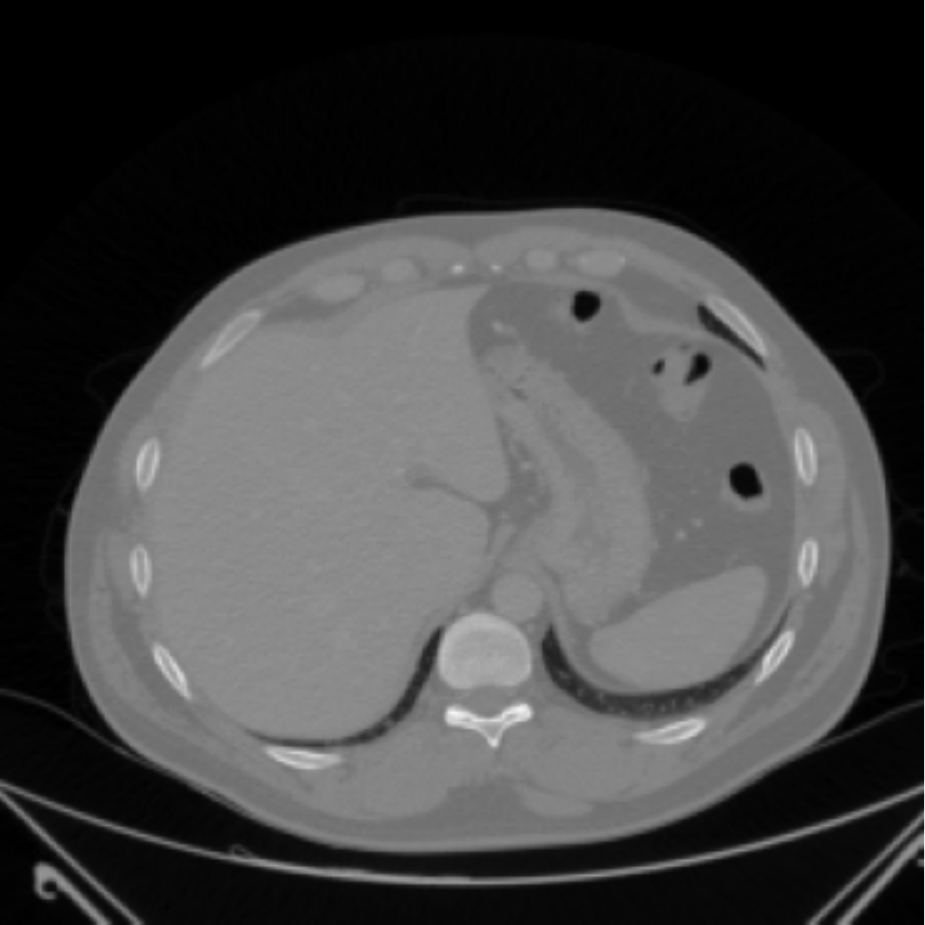}\label{subfig:ct}}} \hspace{1pt}
	\subfloat[MRI T1 in-phase scan.]{{\includegraphics[width=0.24\textwidth]{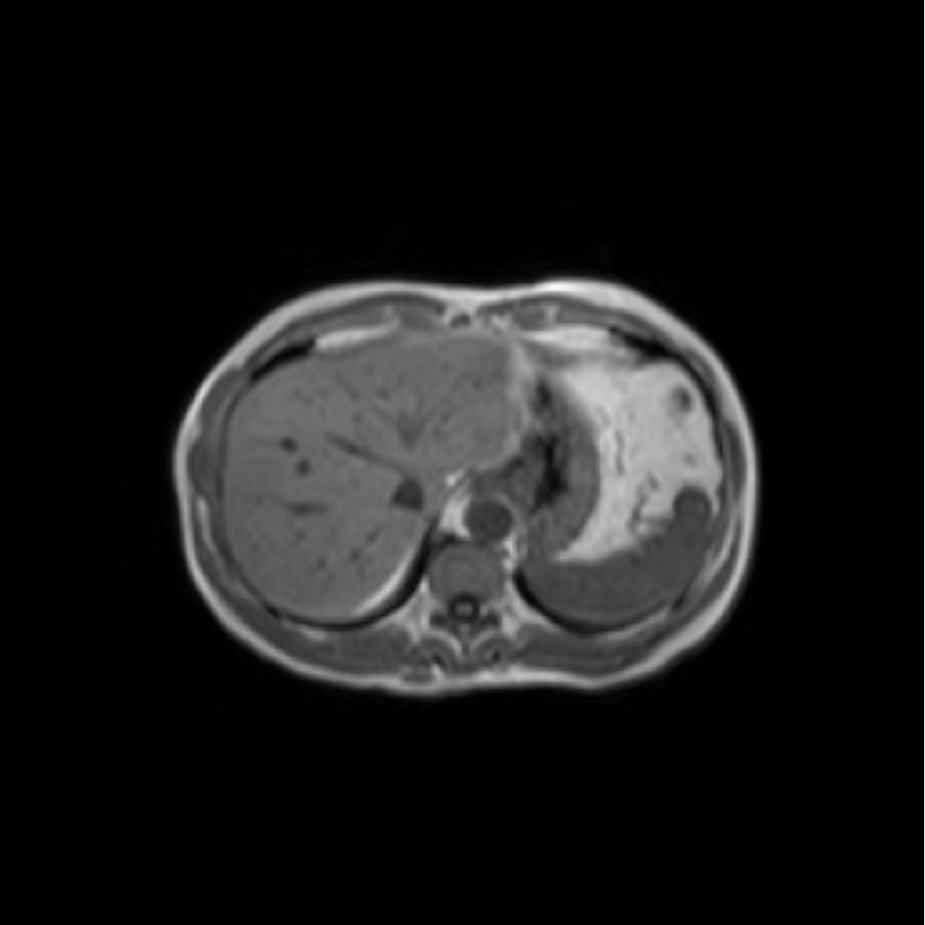}\label{subfig:mri}}}  \hspace{1pt}
	\subfloat[Histograms of entire scans and liver pixels.]{{\includegraphics[width=0.48\textwidth]{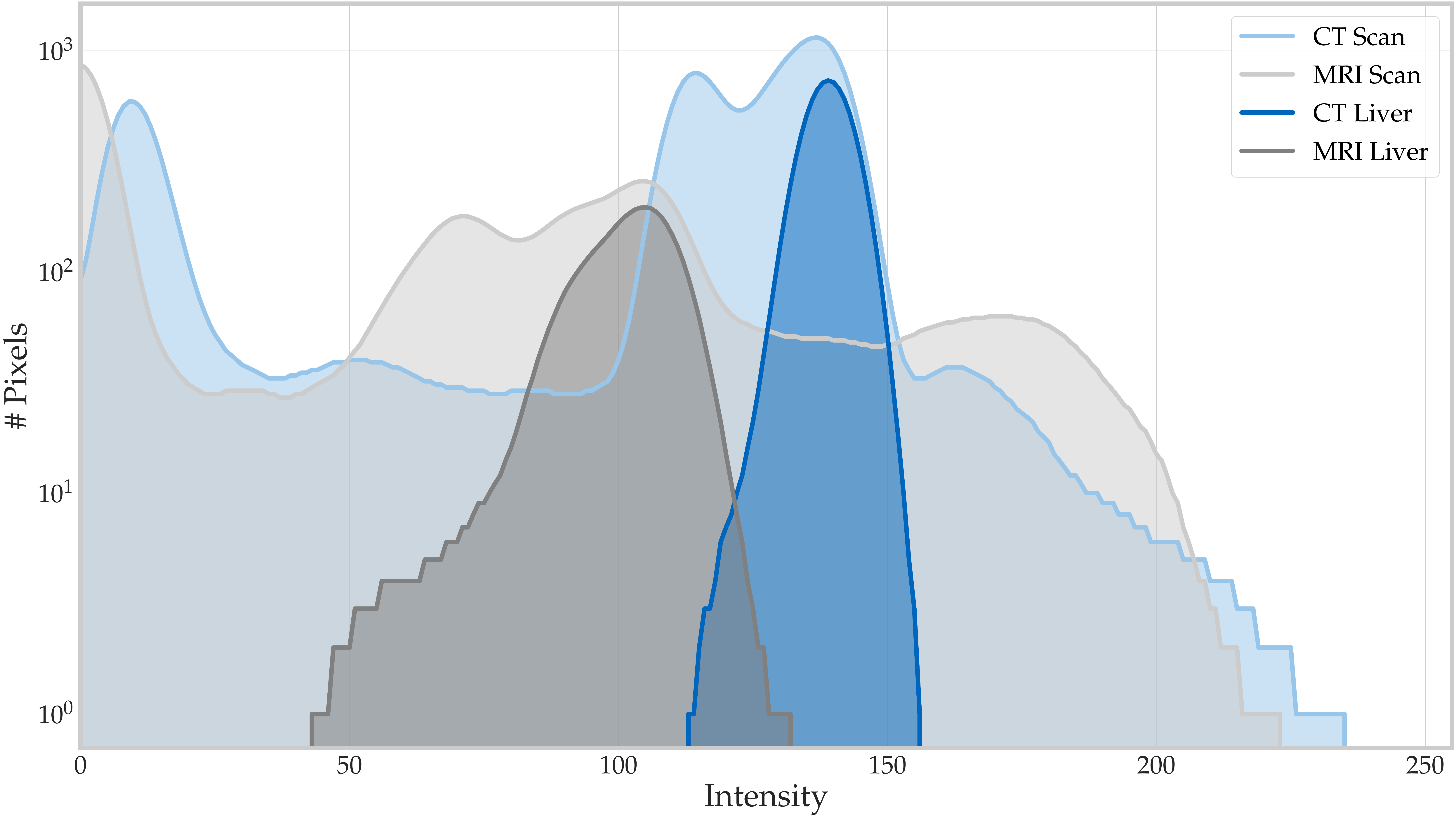}\label{subfig:histogram}}}
	\end{center}
	\end{minipage}
	\caption[Histograms of CT vs. MRI]{CT \subref{subfig:ct} and MRI \subref{subfig:mri} scans from the CHAOS19~\citep{CHAOS19} dataset together with histograms of the whole scans and the liver pixels only \subref{subfig:histogram}. Figure inspired by Yang~et~al.~\citep[Fig.{\,}1]{DADR}.}\label{fig:histogram}
\end{figure}

Due to the high level of data heterogeneity~\citep{DataHeterogeneity} among
the datasets of the clients in the federation, the task of multi-modal 
liver segmentation with FL includes multiple challenges. These challenges 
comprise, but are not limited to: 1) Data from different modalities at the 
clients, i.e. clients with only CT, only MRI, or data from both modalities. 
2) Unbalanced amounts of data at different clients. This may include 
unbalanced amounts of CT and MRI data at clients with data from both 
modalities. 3) The scans are recorded with different scanners and protocols,
and therefore have different image resolutions, sizes and spacing between 
slices. Additionally, scans coming from different imaging modalities have 
different visual appearances (see \Cref{fig:histogram}).

In this work, the aim is to solve these challenges by introducing own approaches. In addition, the 
developed approaches should overcome the limitations of previous methods (see 
\Cref{sec:related-work}). To this end, we formulate a list of key design requirements, which we 
consider when developing our algorithms: 1) Avoid heavy pre-processing steps, e.g. disentangled representations used in the work of Yang~et~al.~\citep{DADR}. 2) No need of having both modalities per client (reflects a real-world scenario). 3) Deal with multi-modal data inside the federation and, if necessary, per client. 4) Handle unbalanced amounts of data per client. 5) Keep the number of model parameters low. 6) Direct applicability of the learned global model to unseen data/clients.

According to the design requirements above, solving the problem of data heterogeneity and multi-modal 
data inside the federation is the main challenge of this work. Due to multiple causes, such as 
different scanners, protocols, resolutions or appearances, a domain shift occurs between both 
modalities~\citep{DADR}. This domain shift can be observed in \Cref{fig:histogram}, which shows the 
histograms of the entire scans and the liver pixels only of a CT and a MRI scan. It can be seen that 
the intensity histograms of the entire scans cover the same area, but are quite dissimilar. The 
histograms of the liver pixels, cover different intensities, which poses a challenge for liver 
segmentation algorithms. For this reason, we decided to use the concept of MN~\citep{ModeNormalization} in our proposed approaches because it is able to capture
different modes in the data, which can be identified in \Cref{fig:histogram}, and to normalize the 
data accordingly.

\subsection{Federated Learning with Modality-Based Normalization}
In this work, we propose two FL algorithms, which leverage the MN 
technique~\citep{ModeNormalization} and normalize data based on 
their modality information. Specifically, MN is a normalization technique 
and an extension of BN~\citep{BatchNormalization}. Inspired by the idea 
that samples inside a dataset/minibatch belong to different latent modes 
with shared common features, 
Deecke et al.~\citep{ModeNormalization} propose to 
separately normalize samples within a minibatch based on these modes, 
instead of normalizing all the samples in the minibatch based on the same 
mean and variance as in BN~\citep{BatchNormalization}. Therefore, MN considers $M$ modes and uses a 
set of gating functions $\{g_m\}_{m=1}^M$, where each function
determines, how likely a sample belongs to mode $m \in \{1, \dots, M\}$.
Due to the functionality of our algorithms, we term the first \fednorm 
and its extension \fednormp. Both algorithms are explained in more detail
in the following.

\subsubsection{FedNorm}\label{sec:fednorm}
While other FL algorithms consider all parameters of all selected clients per round to get averaged at
the server, e.g. \fedavg~\citep{FedAvg}, \fednorm separates between normalization parameters, i.e. all parameters and statistics 
from all MN layers in the neural network, and non-normalization parameters, i.e. all other parameters.
The normalization parameters are further divided into CT and MRI modalities, and the server needs to 
know whether a client contains only CT data or only MRI data. Each modality has its own MN parameters 
and statistics
$\boldsymbol{\theta}^\text{(modality)} := \bigcup_{l \text{ is MN}} \{\boldsymbol{\alpha}^{(l)}, \boldsymbol{\beta}^{(l)}, \overline{\langle \boldsymbol{x} \rangle}^{(l)},\overline{\langle \boldsymbol{x}^2 \rangle}^{(l)}\}$
with $M$ modes, respectively, which contain all parameters ($\boldsymbol{\alpha}$ and 
$\boldsymbol{\beta}$) and statistics ($\overline{\langle \boldsymbol{x} \rangle}$ and 
$\overline{\langle \boldsymbol{x}^2 \rangle}$) from all MN layers in the network. This allows the 
neural network to find latent modes within each modality. When distributing the global model 
parameters to the clients in round $t$, the server sends the non-normalization parameters 
$\boldsymbol{\theta}_t^{}$ together with the normalization parameters
$\boldsymbol{\theta}_t^\text{(CT)}$ for a client with CT data, or $\boldsymbol{\theta}_t^\text{(MRI)}$
for a client with MRI data. Based on the received parameters
$\boldsymbol{\theta} = \big\{\boldsymbol{\theta}_t^{}, \boldsymbol{\theta}_t^{\text{(modality)}}\big\}$,
each client performs local training on its own data for $E$ epochs. In the aggregation step, the 
server averages the non-normalization parameters $\boldsymbol{\theta}_{t+1}^{(k)}$ from all selected 
clients $k \in S_t$ of the current round $t$ in the same way as \fedavg~\citep{FedAvg}.
The CT and MRI normalization parameters and statistics of the MN layers are averaged separately and 
depending on their modality. Additionally, inspired by the \fedavgm algorithm~\citep{FedAvgM}, the aim 
is to avoid oscillations in the parameter updates. Different from \fedavgm, \fednorm uses
$\beta \in (0, 1]$ to interpolate between the old state of the normalization parameters and the new 
averaged normalization parameters. The intention of the separation of normalization parameters into CT
and MRI modalities is to normalize the data within these MN layers to similar states and therefore use
the same non-normalization parameters for both modalities. Hence, the interpolation mechanism is only 
used to update the normalization parameters and is not needed for the non-normalization parameters. A 
detailed description of \fednorm can be found in \Cref{alg:fednorm}.
\begin{algorithm}[ht]
	\begin{algorithmic}[1]
		\SUB{Server executes:}
			\STATE \hspace*{2mm} Initialize $\boldsymbol{\theta}_1^{}, \boldsymbol{\theta}_1^\text{(CT)}, \boldsymbol{\theta}_1^\text{(MRI)}$
			\STATE \hspace*{2mm} \textbf{for} round $t = 1$ to $T$ \textbf{do}
				\STATE \hspace*{4mm} $m \leftarrow \max(C \cdot K, 1)$
				\STATE \hspace*{4mm} $S_t \leftarrow$ (random set of $m$ clients)
				\STATE \hspace*{4mm} \textbf{for} each client $k \in S_t$ \textbf{in parallel do}
					\STATE \hspace*{6mm} \textbf{if} modality(client $k$) == CT \textbf{then}
						\STATE \hspace*{8mm} $\big\{\boldsymbol{\theta}_{t+1}^{(k)}, \boldsymbol{\theta}_{t+1,k}^\text{(CT)}\big\} \leftarrow \text{ClientUpdate}\big(k, \big\{\boldsymbol{\theta}_t^{}, \boldsymbol{\theta}_t^\text{(CT)}\big\}\big)$
					\STATE \hspace*{6mm} \textbf{else if} modality(client $k$) == MRI \textbf{then}
						\STATE \hspace*{8mm} $\big\{\boldsymbol{\theta}_{t+1}^{(k)}, \boldsymbol{\theta}_{t+1,k}^\text{(MRI)}\big\} \leftarrow \text{ClientUpdate}\big(k, \big\{\boldsymbol{\theta}_t^{}, \boldsymbol{\theta}_t^\text{(MRI)}\big\}\big)$
				\STATE \hspace*{4mm} $N \leftarrow \sum_{k \in S_t} N_k$
				\STATE \hspace*{4mm} // \emph{Aggregate non-normalization parameters:}
				\STATE \hspace*{4mm} $\boldsymbol{\theta}_{t+1}^{} \leftarrow \sum_{k \in S_t} \frac{N_k}{N} \boldsymbol{\theta}_{t+1}^{(k)}$
				\STATE \hspace*{4mm} // \emph{Aggregate normalization parameters:}
				\STATE \hspace*{4mm} \textbf{for} modality $\in \{$CT, MRI$\}$ \textbf{do}
					\STATE \hspace*{6mm} $\boldsymbol{\theta}_{t+1}^\text{(modality)} \leftarrow (1 - \beta) \, \boldsymbol{\theta}_t^\text{(modality)} + \beta \, \sum_{k \in S_t} \frac{N_k}{N} \boldsymbol{\theta}_{t+1,k}^\text{(modality)}$
			\vspace{1mm}
		\SUB{ClientUpdate($k, \boldsymbol{\theta}$):}\ \ \ // \emph{Run on client $k$}
			\STATE \hspace*{2mm} $\mathcal{B} \leftarrow$ (split $\mathcal{D}_k$ into batches of size $B$)
			\STATE \hspace*{2mm} \textbf{for} epoch $i = 1$ to $E$ \textbf{do}
				\STATE \hspace*{4mm} \textbf{for} batch $b \in \mathcal{B}$ \textbf{do}
					\STATE \hspace*{6mm} $\boldsymbol{\theta} \leftarrow \boldsymbol{\theta} - \eta \nabla \mathcal{L}(b; \boldsymbol{\theta})$
			\STATE \hspace*{2mm} \textbf{return} $\boldsymbol{\theta}$
	\end{algorithmic}
	\caption[FedNorm]{\fednorm. $K$ is the total number of clients, $k$ is the index of one client, $T$ is the number of FL rounds, $B$ is the local minibatch size,
					$E$ is the number of local epochs, $C \in (0, 1]$ is the fraction of clients selected in each FL round,
					$\eta$ is the local learning rate, $\mathcal{L}$ is a loss function, $\mathcal{D}_k$ is the dataset of client $k$,
					$N_k = \lvert \mathcal{D}_k \rvert$ is the number of samples in dataset $\mathcal{D}_k$,
					and $\beta$ is the interpolation ratio. Algorithm partially adapted from McMahan~et~al.~\cite[Algorithm 1]{FedAvg}.}\label{alg:fednorm}
\end{algorithm}

\subsubsection{FedNorm+}\label{sec:fednormp}
\fednorm separates the MN parameters and statistics by modalities and distributes these parameters to 
the clients based on the modality of their data. As a consequence, it cannot handle mixed-modality 
clients, which contain CT and MRI data, and thus would not be applicable to real-world scenarios. For 
this reason, we propose \fednormp, an extension of \fednorm, which does not separate between 
normalization and non-normalization parameters. Instead, it only uses one set of MN parameters and 
statistics, which is distributed to all selected clients per round in the same way as in 
\fedavg~\citep{FedAvg}. For the MN layers, the number of modes is fixed to $M = 2$ with the intention 
to capture and normalize the samples of each modality (CT and MRI) with one mode, respectively. 
Therefore, the modality information of each scan is handed to the network and forwarded to 
each MN layer. Based on this information, the respective mode is chosen in a hard-coded 
way (cf. original MN technique~\citep{ModeNormalization}):
\begin{itemize}
	\item Mode 1 (CT): Set $g_1(\boldsymbol{x}_n) = 1 \wedge g_2(\boldsymbol{x}_n) = 0$ if
	        modality($\boldsymbol{x}_n$) = CT
	\item Mode 2 (MRI): Set $g_1(\boldsymbol{x}_n) = 0 \wedge g_2(\boldsymbol{x}_n) = 1$ if
	        modality($\boldsymbol{x}_n$) = MRI
\end{itemize}
Consequently, the parameters for each modality are trained on their respective scans. In the 
aggregation step, \fednormp uses the same parameter aggregation as \fedavg~\citep{FedAvg} for all parameters. Due to 
the hard-coded mode assignment, the parameters and statistics for each mode (modality) of a MN layer 
are aggregated separately. Additionally, the interpolation mechanism of \fednorm with momentum $\beta$
is used to update all parameters, instead of only the MN parameters and statistics, to avoid 
oscillations in the updates.
\section{Experiments and Results}\label{sec:experiments-results}
To validate the performance of our proposed approaches, we perform multiple experiments with different
FL settings and datasets. 
We aim to answer the following questions related to data heterogeneity: \textit{i) How well do the models perform when each client only contains data from a single modality? ii) What happens if CT only, MRI only, and mixed-modality (CT + MRI) clients participate in the training?} To address these questions, we carefully designed our Non-IID 1 setting (three CT and three MRI clients) for (i) and our Non-IID 2 setting (two CT, two MRI, and two mixed-modality clients) for (ii).

\subsection{Datasets}\label{sec:datasets}
In our experimental settings, we use multiple publicly available CT and MRI datasets: LiTS17~\citep{LiTS17}, the 3D-IRCADb-01 dataset of 3D-IRCADb~\citep{3DIRCADb} (included in the training set of LiTS17), abdominal scans from the Multi-Atlas dataset~\citep{MultiAtlas}, SLIVER07~\citep{SLIVER07}, and CHAOS19~\citep{CHAOS19}. Furthermore, we use a random subset of 207 T1-weighted MRI scans (patients) from the KORA in-house dataset~\citep{KORA}. Detailed information including 
technical specifications, e.g. numbers of patients (volumes), resolution, modalities, and demographics for all datasets can be found in \Cref{tab:datasets}.
Here, we only use the training sets from LiTS17, Multi-Atlas, SLIVER07, and CHAOS19, for which the 
ground truth liver segmentation masks are provided. For CHAOS19, we compute water-only and fat-only images based on the T1-weighted in-phase and out-phase scans from CHAOS19 using the Dixon technique~\citep{DixonMethod} with the formulas provided in Ma~\citep{DixonTechniques}. 
\begin{table*}[htp]
	\centering
	\caption[Liver Segmentation Datasets]{Technical and demographic information about the liver segmentation datasets used in this work.
											"\# Sites" indicates the number of institutions from which the data was aquired.
											"\# Slices/Patient" gives a range with the minimum and maximum number of slices per patient.
											The age and liver size are reported with mean and standard deviation.
											The resolution (physical size) of a voxel is given in $x$-, $y$-, and $z$-direction,
											where the latter corresponds to the space between neighboring slices.}\label{tab:datasets}
	\resizebox{\textwidth}{!}{%
		\begin{tabular}{l c c c c c c c}
			\toprule
			\textbf{Feature}					& \textbf{LiTS17}~\cite{LiTS17} \textbf{w/o 3D-IRCADb}			& \textbf{3D-IRCADb}~\cite{3DIRCADb}						& \textbf{Multi-Atlas}~\cite{MultiAtlas}					& \textbf{SLIVER07}~\cite{SLIVER07}						& \textbf{CHAOS19}~\cite{CHAOS19} \textbf{CT}		& \textbf{CHAOS19}~\cite{CHAOS19} \textbf{MRI}		& \textbf{KORA}~\cite{KORA}				\\
			\midrule
			\# Sites							& 6										& 1											& Multiple								& Multiple								& 1							& 1							& Multiple					\\
			\# Patients							& 111									& 20										& 30									& 20									& 20						& 20						& 207						\\
			\# Slices							& 55815									& 2823										& 3779									& 4159									& 2874						& 647 (T1), 623 (T2)		& 21379						\\
			\# Slices/Patient					& $[75, 987]$							& $[74, 260]$								& $[85, 198]$							& $[64, 394]$							& $[81, 266]$				& $[26, 50]$				& $[73, 138]$				\\
			Modality							& CT									& CT										& CT									& CT									& CT						& MRI						& MRI						\\
			\multirow{2}{*}{Health Status}		& Most scans with						& 75\% with									& \multirow{2}{*}{n.a.}					& Most scans with						& \multirow{2}{*}{healthy} 	& \multirow{2}{*}{healthy}	& healthy					\\
												& tumors, e.g. HCC						& liver tumors								&										& tumors/metastasis/cysts				&							&							& and (pre-)diabetic		\\
			Age (years)							& n.a.									& $53.6 \pm 13.4$							& n.a.									& n.a.									& n.a. 						& n.a.						& $56.2 \pm 9.3$			\\
			\# Women/Men						& n.a.									& 10/10										& n.a.									& n.a.									& n.a.						& n.a.						& 98/109					\\
			Image Size							& $512 \times 512$						& $512 \times 512$							& $512 \times 512$						& $512 \times 512$						& $512 \times 512$			& \{256, 280, 320\}$^2$ 	& $256 \times 256$			\\
			Scanner								& Multiple								& n.a.										& Multiple								& Multiple								& n.a.						& 1.5T Philips MRI			& 3T Siemens Skyra			\\
			\multirow{3}{*}{Resolution (mm)}	& $[0.56, 0.98] \times$					& $[0.56, 0.87] \times$						& $[0.59, 0.98] \times$					& $[0.58, 0.81] \times$					& $[0.57, 0.79] \times$		& $[1.36, 2.03] \times$		& $[1.7, 1.7] \times$		\\
												& $[0.56, 0.98] \times$					& $[0.56, 0.87] \times$						& $[0.59, 0.98] \times$					& $[0.58, 0.81] \times$					& $[0.57, 0.79] \times$		& $[1.36, 2.03] \times$		& $[1.67, 1.74] \times$		\\
												& $\,[0.7, 5.0]$						& $\,[1.0, 4.0]$							& $\,[2.5, 5.0]$						& $\,[0.7, 5.0]$						& $\,[1.0, 2.0]$			& $\,[5.5, 9.0]$			& $\,[1.67, 1.74]$			\\
			Liver Size (cm$^3$)					& $1694.398 \pm 406.093$				& $1583.721 \pm 272.122$					& $1738.907 \pm 446.85$					& $1710.61 \pm 434.628$					& $1592.109 \pm 281.597$	& $1609.635 \pm 500.653$	& $1608.312 \pm 376.634$	\\
			\bottomrule
		\end{tabular}
	}
\end{table*}

We cropped (left and right in axial view) the scans from the KORA dataset~\citep{KORA} and padded them with 
zeros (top and bottom in axial view) to obtain images of size $256 \times 256$ pixels. Furthermore, we 
removed many, but not all, slices that do not contain the liver. Consequently, the models can be trained to
predict a segmentation for the liver only on images on which it is present.

The scans in the other datasets have varying image sizes (see \Cref{tab:datasets}) and intensity value
ranges. Therefore, the scans are first resized to a size of $256 \times 256$ pixels and the
intensity values of the 2D slices are normalized based on the single scalar mean and standard 
deviation values that are obtained by taking the mean and standard deviation of all intensities per
slice and taking the mean over all slices from all volumes in the dataset. Resizing the CT scans from 
$512 \times 512$ pixels to $256 \times 256$ pixels also has the effect of doubling their image 
resolution ($x$-$y$-plane), resulting in approximately the same image resolution as the MRI scans. 
Finally, the normalized values are clipped to $[-3, 3]$ as proposed by 
Stan~et~al.~\citep{Preprocessing}.

\subsection{Experiments}\label{sec:experiments}
\begin{table*}[ht]
	\caption[Federated Learning Settings]{Experimental FL settings based on data from the
	                                        Multi-Atlas~\cite{MultiAtlas}, 3D-IRCADb~\cite{3DIRCADb},
											SLIVER07~\cite{SLIVER07}, CHAOS19~\cite{CHAOS19}, LiTS17~\cite{LiTS17}, and KORA~\cite{KORA} datasets.
											The last three columns indicate the splitting of the volumes (patients)
											in the datasets of the clients into training, validation and testing parts. For each client dataset, a standalone
											local model is trained as an upper bound baseline. Furthermore, a centralized model for each setting is trained on
											the collection of training data (data lake) from all clients in the setting.}\label{tab:fl-settings}
	\centering
	\resizebox{\textwidth}{!}{%
		\begin{tabular}{c c l c c c c}
			\toprule
			\multirow{2}{*}{\textbf{Setting}}	& \multirow{2}{*}{\textbf{Models}}							& \multirow{2}{*}{\textbf{Datasets}}	& \multirow{2}{*}{\textbf{Modality}}	& \textbf{\# Training}	& \textbf{\# Validation}	& \textbf{\# Testing}	\\
												&															&										&										& \textbf{Volumes}		& \textbf{Volumes}			& \textbf{Volumes}		\\
			\midrule
			\multirow{7}{*}{Non-IID 1}			& \multirow{6}{*}{\shortstack[c]{Clients/\\Local Models}}	& Multi-Atlas							& CT									& 8						& 4 						& 18 					\\
												&															& 3D-IRCADb								& CT									& 8						& 4 						& 8						\\
												&															& SLIVER07								& CT									& 8						& 4 						& 8						\\
												&															& CHAOS19 MR T1 In-Phase				& MRI									& 8						& 4 						& 8						\\
												&															& CHAOS19 MR T1 Out-Phase				& MRI									& 8						& 4 						& 8						\\
												&															& CHAOS19 MR T1 Water					& MRI									& 8						& 4 						& 8						\\
												  \cdashline{2-7}
												& Centralized Model											& All Clients							& CT + MRI								& 48					& 24						& 58					\\
			\midrule
			\multirow{7}{*}{Non-IID 2}			& \multirow{6}{*}{\shortstack[c]{Clients/\\Local Models}}	& Multi-Atlas							& CT									& 8						& 4	 						& 18					\\
												&															& 3D-IRCADb								& CT									& 8						& 4							& 8						\\
												&															& CHAOS19 MR T1 Water						& MRI									& 8						& 4							& 8						\\
												&															& KORA T1 Water (1)						& MRI									& 8						& 4							& 8						\\
												&															& SLIVER07 + KORA T1 Water (2)			& CT + MRI								& 16 (8 + 8)			& 8 (4 + 4)					& 16 (8 + 8)			\\
												&															& LiTS17 Site 2 + KORA T1 Water (3)		& CT + MRI								& 14 (6 + 8)			& 7 (3 + 4)					& 14 (6 + 8)			\\
												  \cdashline{2-7}
												& Centralized Model											& All Clients							& CT + MRI								& 62					& 31						& 72					\\
			\bottomrule
		\end{tabular}
	}
\end{table*}
We use two different experimental FL settings to train, validate and test our proposed approaches, 
\fednorm (\Cref{sec:fednorm}) and \fednormp (\Cref{sec:fednormp}), and compare them against baselines 
and other state-of-the-art methods. Specifically, we design two non-IID settings, which contain clients with datasets from both modalities and which cover different scenarios:
The Non-IID 1 setting contains a balanced number of CT and MRI clients, where the dataset of each client only contains data from one modality. The Non-IID 2 setting includes CT, MRI 
and mixed-modality (CT + MRI) clients. However, the Non-IID 1 setting is biased in the MRI clients 
because all of them show the same patients (shapes), but in different MRI sequences (appearances). 
Specifically, the designed FL settings contain an increasing amount of data 
heterogeneity~\citep{DataHeterogeneity} from the Non-IID 1 to the Non-IID 2 setting, due to the mixed-modality clients. Detailed information about both FL settings is provided in \Cref{tab:fl-settings}. Here, the clients correspond to different
hospitals, each with its own training, validation and testing sets from one (or two) dataset(s), and 
thus with specific scanner types and protocols per client, which were used to record the medical 
images. For the client datasets, we mainly use a splitting of 40\% - 20\% - 40\% (instead of the more 
common 80\% - 10\% - 10\%) because FL aggregates the client models, which in turn is similar to 
training on an aggregation of the training sets from the clients. Another reason for this partitioning
of the data is to have as few patients as possible per client for training, to reflect a real-world 
scenario with rarely labeled volumes in each hospital, and to have as many volumes as possible for
testing. To evaluate the performance of the models, we use the Dice 
coefficient~\citep{Dice}.

\subsubsection{Model Selection}\label{sec:model-selection}
Before comparing our proposed approaches to other methods, we select winners for \fednorm 
(\Cref{sec:fednorm}) and \fednormp (\Cref{sec:fednormp}) by tuning their hyper-parameters. Here, we 
follow the proposal of Sheller~et~al.~\citep{MedicalFL} and select a winner based on the 
highest global validation score reported during training. According to
Sheller~et~al.~\citep{MedicalFL}, this global validation score is computed by aggregating the 
local validation scores of all clients in the federation, which validate the global model at the 
beginning of each round on their local validation set before starting their training. However, we do 
not include all clients for training in each round, but we compute the global validation score based 
on the local validation scores of all clients in the federation. Therefore, we compute the mean 
(instead of a weighted average based on the amount of client data) of local validation scores to 
obtain the global validation score. Consequently, the validation score of each client contributes 
equally to the global score, which then measures the overall performance on the data of all clients.

\paragraph{FedNorm}
For \fednorm, we focus on the number of modes $M$ per modality and the interpolation ratio $\beta$, 
which is used for aggregating the averaged new and old normalization parameters, 
as hyper-parameters. For this reason, we investigate the influence of $M \in \{1, 2, 3, 4\}$ and 
$\beta \in \{1.0, 0.9, 0.5, 0.2\}$, where the old state of the normalization parameters is ignored with
$\beta = 1.0$, and $\beta = 0.2$ only slightly changes the old normalization parameters in the direction of the 
averaged new ones. To this end, we train \fednorm with each combination of $M$ and $\beta$ for 100 FL 
rounds with two clients per round and $E = 1$ local epoch on the Non-IID 1 setting (see 
\Cref{tab:fl-settings}) because it is the only setting which \fednorm can handle. The highest global validation Dice 
scores for all combinations within these 100 FL rounds of training are reported in 
\Cref{tab:model-selection}.
Based on these scores, we select \fednorm with $M = 2$ modes and $\beta = 0.9$ as the winner and use 
it in our further experiments.
\begin{table}[ht]
	\caption[Model Selection: FedNorm and FedNorm+]{Highest global validation Dice scores for \fednorm and \fednormp with different interpolation ratios $\beta$ and varying numbers of
										modes $M$ per modality reported within 100 FL rounds of training on the Non-IID 1 (\fednorm) and Non-IID 2 (\fednormp) settings. The best global validation Dice score per method is highlighted in bold.}\label{tab:model-selection}
	\centering
		\begin{tabular}{l c c c c c}
			\toprule
            \multirow{2}{*}{Method}	    & \multirow{2}{*}{\shortstack{\textbf{\# Modes}\\($\boldsymbol{M}$)}} & \multicolumn{4}{c}{\textbf{Interpolation Ratio} $\boldsymbol{(\beta)}$}   \\
            							& 						                                & \textbf{1.0}	& \textbf{0.9}	& \textbf{0.5}	& \textbf{0.2}  \\
			\midrule
            \multirow{4}{*}{FedNorm}    & 1									                    & 0.889			& 0.880			& 0.880			& 0.860	        \\
            			                & 2									                    & 0.892			& \textbf{0.893}& 0.889         & 0.863	        \\
            			                & 3									                    & 0.888			& 0.889			& 0.882			& 0.868	        \\
            			                & 4									                    & 0.889			& 0.885			& 0.880			& 0.840	        \\
            \midrule
            FedNorm+                    & 2                                                     & 0.876         & 0.877         & \textbf{0.879}& 0.831            \\
			\bottomrule
		\end{tabular}
\end{table}

\paragraph{FedNorm+}
Since the number of modes is fixed to $M = 2$ for \fednormp (see \Cref{sec:fednormp}), we only focus 
on the interpolation ratio $\beta$ as a hyper-parameter. Similar to \fednorm, we investigate the influence of 
$\beta \in \{1.0, 0.9, 0.5, 0.2\}$. Therefore, we train \fednormp for 100 FL rounds with two clients per round
and $E = 1$ local epoch on the Non-IID 2 setting (see \Cref{tab:fl-settings}) because it is our FL setting 
with the highest degree of data heterogeneity~\citep{DataHeterogeneity}.

The highest global validation Dice scores within the 100 FL rounds of training are reported in 
\Cref{tab:model-selection}.
Based on these values, we select \fednormp with $\beta = 0.5$ as the winner and use it in our further 
experiments.

\subsubsection{Implementation Details}
Due to the great success of the U-Net~\citep{UNet} neural network 
architecture in biomedical image segmentation, we employ a 
shallower version of the original network for the task 
of liver segmentation on 2D slices in this work.
Here, we reduce the feature map channels of the original 
U-Net~\citep{UNet} from $\{64, 128, 256, 512, 1024\}$ to
$\{8, 16, 32, 64\}$, resulting in ${\sim}136$K instead of ${\sim}36$M 
parameters. Consequently, this modified version of the 
U-Net meets the desired requirement to keep the number of model 
parameters low (see \Cref{sec:challenges}). The final ($1 \times 1$)
convolution of our modified U-Net computes an output with only one 
channel, followed by the sigmoid function~\citep[Equation 4.59]{Bishop}
to compute a pixel-wise pseudo-probability for each pixel, which 
tells how likely it belongs to the liver.

Furthermore, zero-padding of size one is used to obtain the same 
output size as the input size for the $3 \times 3$ convolutions.

For the state-of-the-art FL algorithms, \fedavg~\citep{FedAvg}, \fedavgm~\citep{FedAvgM}, 
\fedvc~\citep{FedVC}, \silobn~\citep{SiloBN}, and \fedbn~\citep{FedBN}, we use our own re-implementation 
based on the information in these works to better fit our needs and code framework. For the 
experiments with \fedavg, \fedavgm, \fedvc, and the local models (see \Cref{tab:fl-settings}), we use our modified version of the U-Net. For \fedavgm, we set the momentum parameter to $\beta = 0.6$ in 
both settings.
The number of local updates $S$ in \fedvc is limited to the minimum number of training samples 
(slices) among all $K$ clients in the federation divided by the batch size $B$ (cf. Hsu~et~al.~\citep{FedVC}),
$S = \lfloor \min\{N_1, \dots, N_K\} / B \rfloor$,
where $N_k$ is the training dataset size of client $k$.
For \silobn and \fedbn, BN~\citep{BatchNormalization} layers are inserted between each convolution 
layer and Rectified Linear Unit (ReLU) into our modified version of the U-Net, except for the last convolution layer. For \fednorm (\Cref{sec:fednorm}), \fednormp (\Cref{sec:fednormp}), and 
the centralized models (see \Cref{tab:fl-settings}), MN~\citep{ModeNormalization} layers instead
of BN layers are inserted at the same position into our modified U-Net. Here, we use MN with $M = 2$ 
for all \fednorm experiments (see \Cref{sec:model-selection}). For the centralized models, we use MN 
with $M = 4$ modes in the modified U-Net of the Non-IID 1 setting (because \fednorm has $M$ modes for each of the two modalities) and MN with $M = 2$ modes for the Non-IID 2 setting to guarantee a fair comparison to \fednorm and \fednormp (see \Cref{sec:model-selection}). We take the implementation of MN from publicly available
code~\footnote{\url{https://github.com/ldeecke/mn-torch} (Accessed: 2021-08-28)}
from Deecke~et~al.~\citep{ModeNormalization}.

The training of all local and centralized models, and at each 
client for the FL algorithms was performed with the Adam 
optimizer~\citep{Adam} with a learning rate of $\eta = 0.001$ and a batch 
size of $B = 12$. All local and centralized models are trained for 50 epochs
with possible early stopping after a patience of five epochs.
We train the models on both FL settings (see \Cref{tab:fl-settings}) for
$T = 100$ FL rounds. In each round, two clients are sampled randomly, but
fixed to be the same for all FL algorithms. Inspired by the work of 
Zhao~et~al.~\citep{FLNonIID}, we train a neural network for $E = 1$ local 
epoch at the client side to prevent a large divergence of the federated 
global model parameters from parameters that the centralized model would 
learn. Our code is made publicly available~\footnote{\url{https://github.com/albarqounilab/FedNorm}}.

All models are trained in a supervised fashion to minimize a total 
loss. Inspired by Isensee~et~al.~\citep{nnUNet}, this total loss 
is composed of the Dice loss~\citep{VNet} and the cross-entropy 
loss~\citep{Bishop}. Specifically, we use the binary cross-entropy 
(BCE) loss \citep[cf. Equation 4.90]{Bishop} $\mathcal{L}_\text{BCE}$
for two classes.
Hence, the total loss is
$\mathcal{L} = \mathcal{L}_\text{Dice} + \mathcal{L}_\text{BCE}$
(cf. Isensee~et~al.~\citep[Equation 1]{nnUNet}). However, different from 
the Dice loss proposed by Milletari~et~al.~\citep{VNet}, we use the 
non-binary pseudo-probabilities $p_{ij} \in [0, 1]$ instead of binary segmentations to compute the 
loss value (cf. Kodym~et~al.~\citep{DiceLoss}).
Nevertheless, binary segmentation results are used to compute the Dice coefficient~\citep{Dice} values in the evaluation, where we round the non-binary predictions $p_{ij}$ of the network to binary values: $\hat{y}_{ij} = 0$, if $p_{ij} \in [0, 0.5]$, and $\hat{y}_{ij} = 1$, if $p_{ij} \in (0.5, 1]$.

\subsubsection{Comparison with Baselines and State of the Art}\label{sec:comparison}
We compare \fednorm (\Cref{sec:fednorm}) and \fednormp (\Cref{sec:fednormp}) to different baselines and 
state-of-the-art FL algorithms in both FL settings (see \Cref{tab:fl-settings}). As baselines, we consider local 
models and centralized models. The local models are trained on the training set of each client, individually. 
Consequently, the performance of these models can be seen as upper bounds for each client dataset. Additionally, per setting, we train one centralized model on the collection of training data (data lake) from all clients in
the setting. Specifically, \Cref{tab:fl-settings} gives the number of training, validation and testing volumes (patients) for all FL clients, local and centralized models in each setting.

As state-of-the-art FL algorithms, we choose the base FL algorithm, \fedavg~\citep{FedAvg}, and different 
approaches, which tackle the problem of data heterogeneity~\citep{DataHeterogeneity} in FL, to compare against \fednorm and \fednormp: \fedavgm~\citep{FedAvgM}, \fedvc~\citep{FedVC}, \silobn~\citep{SiloBN}, and \fedbn~\citep{FedBN}.
For all FL algorithms, the global model from the server aggregation of the last FL round is sent back to each client and is tested on the test set of each client without any refinement. To compare the different methods, we measure the performance of all models on the test
sets of all clients in both settings (\Cref{tab:fl-settings}) with the Dice coefficient~\citep{Dice}. The resulting performance values for both settings are shown in
\Cref{tab:comparison} and will be investigated in further detail in the following. To indicate the difference in performance between the federated and centralized models, and the local models, the relative improvement (RI)~\citep{RelativeImprovement} is computed as
$\text{RI} = \frac{x_2 - x_1}{x_1} \cdot 100 \%$~\citep[cf. Equation 9]{RelativeImprovement},
which gives the improvement (in percent) of performance score $x_2$ over score $x_1$, where both are measured 
with the same metric. Additionally, the unpaired two-sided t-test is performed
on the Dice per patient results of the local model and each federated or centralized model, respectively, where a 
$p$-value $\leq 0.05$ indicates statistical significance.

In \Cref{tab:comparison} it can be seen that the federated models consistently (except for CHAOS19 MR T1 
Out-Phase and Water) outperform the local models (positive RI). Furthermore, in many cases, the 
performances of some federated models are similar or even better than the ones from the centralized models, which use more volumes, coming from both modalities, directly for model training. 
Additionally, it can be seen that \fednorm and \fednormp show the best performance per client in many cases
and outperform the centralized models on most of the client's test sets. Furthermore, it has to be noted 
that the model from \fednormp is the only model that can outperform the local model on the
CHAOS19 MR T1 Water client in the Non-IID 2 setting, and by a large margin.
\begin{table*}[ht]
	\caption[Comparison: Non-IID 1 and 2 Settings]{Comparison on the Non-IID 1 and 2 settings. The numbers give the performances of the local models, the centralized model, and the
											federated global models (without refinement at the clients) on the test set of each client. The performances are reported with Dice per patient and are given by mean (median) and standard deviation.
											Additionally, the RI is computed for the mean Dice per patient of the models from the FL algorithms and the centralized
											models over the local model performances. The best score from the federated models per setting
											and client is highlighted in bold. Additionally, the unpaired two-sided t-test was performed using the Dice per patient scores from the local model
											against the results from each federated model and the centralized model per client. A star ($^\ast$) behind respective RI value indicates statistical
											significance with a $p$-value $\leq 0.05$.}\label{tab:comparison}
	\centering
	\begin{adjustbox}{max width=\textwidth, max height=0.45\textheight}
		\begin{tabular}{l c c c c c c}
 			\toprule
			\multirow{3.5}{*}{\textbf{Methods}}									& \multicolumn{3}{c}{\textbf{Non-IID 1}} 																												& \multicolumn{3}{c}{\textbf{Non-IID 2}}   	\\
			\cmidrule(r){2-4} \cmidrule(l){5-7}
																				&	\textbf{Client}	    											& \textbf{Dice per}                         & \textbf{RI} $\boldsymbol{\uparrow}$   & \textbf{Client}																	& \textbf{Dice per}                      	& \textbf{RI} $\boldsymbol{\uparrow}$   \\
																				&	\textbf{(Modality)} 											& \textbf{Patient} $\boldsymbol{\uparrow}$	& \textbf{[\%]} 			            & \textbf{(Modality)}																& \textbf{Patient} $\boldsymbol{\uparrow}$	& \textbf{[\%]}                    		\\
			\midrule
			Local																& \multirow{9}{*}{\shortstack{Multi-Atlas\\(CT)}}					& $0.850\,(0.862) \pm 0.050$  				& -										& \multirow{9}{*}{\shortstack{Multi-Atlas\\(CT)}}									& $0.850\,(0.862) \pm 0.050$				& -										\\
			\cdashline{1-1}														\cdashline{3-4}\cdashline{6-7}
			FedAvg~\cite{FedAvg}												&																	& $0.883\,(0.899) \pm 0.071$  				& 3.9 									&																					& $0.906\,(0.907) \pm 0.042$				& 6.6$^\ast$						  	\\
			FedAvgM~\cite{FedAvgM}												&																	& $0.575\,(0.541) \pm 0.150$  				& -32.4$^\ast$							&																					& -	                						& -									  	\\
			FedVC~\cite{FedVC}													&																	& $0.884\,(0.885) \pm 0.036$  				& 4.0$^\ast$							&																					& $0.848\,(0.843) \pm 0.056$				& -0.2								  	\\
			SiloBN~\cite{SiloBN}												&																	& $0.870\,(0.909) \pm 0.117$  				& 2.4 									&																					& $0.894\,(0.895) \pm 0.045$				& 5.2$^\ast$						  	\\
			FedBN~\cite{FedBN}													&																	& $0.900\,(0.921) \pm 0.053$  				& 5.9$^\ast$							&																					& $\boldsymbol{0.916\,(0.919) \pm 0.034}$	& \textbf{7.8}$^\ast$				  	\\
			FedNorm																&																	& $\boldsymbol{0.921\,(0.932) \pm 0.033}$	& \textbf{8.4}$^\ast$					&																					& -                 						& -									 	\\
			FedNorm+															&																	& $0.891\,(0.898) \pm 0.034$  				& 4.8$^\ast$							&																					& $\boldsymbol{0.916\,(0.928) \pm 0.030}$	& \textbf{7.8}$^\ast$				 	\\
			\cdashline{1-1}														\cdashline{3-4}\cdashline{6-7}
			Centralized															&																	& $0.921\,(0.934) \pm 0.033$  				& 8.4$^\ast$							&																					& $0.894\,(0.914) \pm 0.061$				& 5.2$^\ast$						  	\\
 			\midrule
 			Local																& \multirow{9}{*}{\shortstack{3D-IRCADb\\(CT)}}						& $0.918\,(0.926) \pm 0.029$  				& -     								& \multirow{9}{*}{\shortstack{3D-IRCADb\\(CT)}}										& $0.918\,(0.926) \pm 0.029$				& -										\\
			\cdashline{1-1}														\cdashline{3-4}\cdashline{6-7}
			FedAvg~\cite{FedAvg}												&																	& $0.940\,(0.942) \pm 0.018$  				& 2.4 									&																					& $0.933\,(0.946) \pm 0.037$				& 1.6								  	\\
			FedAvgM~\cite{FedAvgM}												&																	& $0.874\,(0.888) \pm 0.036$  				& -4.8$^\ast$							&																					& -                 						& -									  	\\
			FedVC~\cite{FedVC}													&																	& $0.929\,(0.933) \pm 0.022$  				& 1.2 									&																					& $0.916\,(0.916) \pm 0.024$				& -0.2								  	\\
			SiloBN~\cite{SiloBN}												&																	& $0.954\,(0.958) \pm 0.014$  				& 3.9$^\ast$							&																					& $\boldsymbol{0.961\,(0.967) \pm 0.011}$	& \textbf{4.7}$^\ast$				  	\\
			FedBN~\cite{FedBN}													&																	& $0.954\,(0.954) \pm 0.015$  				& 3.9$^\ast$							&																					& $0.960\,(0.963) \pm 0.011$				& 4.6$^\ast$						  	\\
			FedNorm																&																	& $\boldsymbol{0.956\,(0.959) \pm 0.017}$	& \textbf{4.1}$^\ast$					&																					& -                 						& -									 	\\
			FedNorm+															&																	& $0.942\,(0.942) \pm 0.016$  				& 2.6 									&																					& $0.955\,(0.956) \pm 0.016$				& 4.0$^\ast$						 	\\
			\cdashline{1-1}														\cdashline{3-4}\cdashline{6-7}
			Centralized															&																	& $0.953\,(0.957) \pm 0.012$  				& 3.8$^\ast$							&																					& $0.938\,(0.945) \pm 0.028$				& 2.2								  	\\
 			\midrule
 			Local																& \multirow{9}{*}{\shortstack{SLIVER07\\(CT)}}						& $0.888\,(0.887) \pm 0.062$  				& -    									& \multirow{9}{*}{\shortstack{CHAOS19 MR\\T1 Water\\(MRI)}}							& $0.718\,(0.784) \pm 0.209$				& -										\\
			\cdashline{1-1}														\cdashline{3-4}\cdashline{6-7}
			FedAvg~\cite{FedAvg}												&																	& $0.888\,(0.924) \pm 0.135$  				& 0.0 									&																					& $0.421\,(0.472) \pm 0.339$				& -41.4								  	\\
			FedAvgM~\cite{FedAvgM}												&																	& $0.761\,(0.785) \pm 0.122$  				& -14.3$^\ast$							&																					& -                 						& -									  	\\
			FedVC~\cite{FedVC}													&																	& $0.905\,(0.926) \pm 0.079$  				& 1.9 									&																					& $0.424\,(0.493) \pm 0.277$				& -40.9$^\ast$						  	\\
			SiloBN~\cite{SiloBN}												&																	& $\boldsymbol{0.924\,(0.944) \pm 0.070}$	& \textbf{4.1}							&																					& $0.514\,(0.498) \pm 0.233$				& -28.4								  	\\
			FedBN~\cite{FedBN}													&																	& $0.906\,(0.928) \pm 0.064$  				& 2.0 									&																					& $0.459\,(0.512) \pm 0.319$				& -36.1								  	\\
			FedNorm																&																	& $0.914\,(0.940) \pm 0.084$  				& 2.9 									&																					& -                 						& -									 	\\
			FedNorm+															&																	& $0.883\,(0.903) \pm 0.116$  				& -0.6 									&																					& $\boldsymbol{0.759\,(0.754) \pm 0.065}$	& \textbf{5.7}						 	\\
			\cdashline{1-1}														\cdashline{3-4}\cdashline{6-7}
			Centralized															&																	& $0.934\,(0.943) \pm 0.057$  				& 5.2 									&																					& $0.893\,(0.879) \pm 0.082$				& 24.4								  	\\
			\midrule
			Local																& \multirow{9}{*}{\shortstack{CHAOS19 MR\\T1 In-Phase\\(MRI)}}		& $0.254\,(0.278) \pm 0.168$				& -   									& \multirow{9}{*}{\shortstack{KORA\\T1 Water (1)\\(MRI)}}							& $0.865\,(0.870) \pm 0.043$				& -										\\
			\cdashline{1-1}														\cdashline{3-4}\cdashline{6-7}
			FedAvg~\cite{FedAvg}												&																	& $0.460\,(0.511) \pm 0.233$  				& 81.1 									&																					& $0.930\,(0.935) \pm 0.030$				& 7.5$^\ast$						  	\\
			FedAvgM~\cite{FedAvgM}												&																	& $0.579\,(0.661) \pm 0.199$  				& 128.0$^\ast$							&																					& -                 						& -									  	\\
			FedVC~\cite{FedVC}													&																	& $0.228\,(0.070) \pm 0.301$  				& -10.2 								&																					& $0.893\,(0.896) \pm 0.044$				& 3.2								  	\\
			SiloBN~\cite{SiloBN}												&																	& $0.656\,(0.678) \pm 0.185$  				& 158.3$^\ast$							&																					& $0.932\,(0.933) \pm 0.026$				& 7.7$^\ast$						  	\\
			FedBN~\cite{FedBN}													&																	& $0.386\,(0.485) \pm 0.333$  				& 52.0 									&																					& $\boldsymbol{0.941\,(0.945) \pm 0.021}$	& \textbf{8.8}$^\ast$				  	\\
			FedNorm																&																	& $\boldsymbol{0.795\,(0.851) \pm 0.170}$	& \textbf{213.0}$^\ast$					&																					& -                 						& -									 	\\
			FedNorm+															&																	& $0.653\,(0.803) \pm 0.303$  				& 157.1$^\ast$							&																					& $0.938\,(0.941) \pm 0.022$				& 8.4$^\ast$						 	\\
			\cdashline{1-1}														\cdashline{3-4}\cdashline{6-7}
			Centralized															&																	& $0.688\,(0.869) \pm 0.369$  				& 170.9$^\ast$							&																					& $0.936\,(0.939) \pm 0.024$				& 8.2$^\ast$						  	\\
			\midrule
            Local																& \multirow{9}{*}{\shortstack{CHAOS19 MR\\T1 Out-Phase\\(MRI)}}		& $0.704\,(0.735) \pm 0.182$  				& -   									& \multirow{9}{*}{\shortstack{SLIVER07\\+ KORA T1\\Water (2)\\(CT + MRI)}}			& $0.703\,(0.872) \pm 0.348$				& -										\\
			\cdashline{1-1}														\cdashline{3-4}\cdashline{6-7}
			FedAvg~\cite{FedAvg}												&																	& $0.443\,(0.477) \pm 0.237$  				& -37.1$^\ast$							&																					& $0.793\,(0.928) \pm 0.282$				& 12.8								  	\\
			FedAvgM~\cite{FedAvgM}												&																	& $0.590\,(0.672) \pm 0.248$  				& -16.2 								&																					& -                 						& -									  	\\
			FedVC~\cite{FedVC}													&																	& $0.296\,(0.285) \pm 0.307$  				& -58.0$^\ast$							&																					& $0.733\,(0.886) \pm 0.333$				& 4.3								  	\\
			SiloBN~\cite{SiloBN}												&																	& $0.735\,(0.713) \pm 0.137$  				& 4.4 									&																					& $\boldsymbol{0.823\,(0.923) \pm 0.214}$	& \textbf{17.1}						  	\\
			FedBN~\cite{FedBN}													&																	& $0.444\,(0.460) \pm 0.358$  				& -36.9 								&																					& $0.784\,(0.927) \pm 0.282$				& 11.5								  	\\
			FedNorm																&																	& $\boldsymbol{0.777\,(0.786) \pm 0.090}$	& \textbf{10.4} 						&																					& -                 						& -									 	\\
			FedNorm+															&																	& $0.661\,(0.752) \pm 0.271$  				& -6.1 									&																					& $0.755\,(0.933) \pm 0.347$				& 7.4								 	\\
			\cdashline{1-1}														\cdashline{3-4}\cdashline{6-7}
			Centralized															&																	& $0.716\,(0.867) \pm 0.318$  				& 1.7 									&																					& $0.844\,(0.925) \pm 0.165$				& 20.1								  	\\
            \midrule
            Local																& \multirow{9}{*}{\shortstack{CHAOS19 MR\\T1 Water\\(MRI)}}			& $0.718\,(0.784) \pm 0.209$  				& -    									& \multirow{9}{*}{\shortstack{LiTS17\\Site 2\\+ KORA T1\\Water (3)\\(CT + MRI)}}	& $0.885\,(0.918) \pm 0.089$				& -										\\
			\cdashline{1-1}														\cdashline{3-4}\cdashline{6-7}
			FedAvg~\cite{FedAvg}												&																	& $0.481\,(0.578) \pm 0.280$  				& -33.0 								&																					& $0.932\,(0.953) \pm 0.050$				& 5.3								  	\\
			FedAvgM~\cite{FedAvgM}												&																	& $0.604\,(0.672) \pm 0.240$  				& -15.9 								&																					& -                 						& -									  	\\
			FedVC~\cite{FedVC}													&																	& $0.291\,(0.248) \pm 0.311$  				& -59.5$^\ast$							&																					& $0.882\,(0.909) \pm 0.089$				& -0.3								  	\\
			SiloBN~\cite{SiloBN}												&																	& $0.683\,(0.685) \pm 0.164$  				& -4.9 									&																					& $0.918\,(0.942) \pm 0.070$				& 3.7								  	\\
			FedBN~\cite{FedBN}													&																	& $0.467\,(0.543) \pm 0.342$  				& -35.0 								&																					& $0.810\,(0.915) \pm 0.174$				& -8.5								  	\\
			FedNorm																&																	& $\boldsymbol{0.807\,(0.841) \pm 0.126}$	& \textbf{12.4} 						&																					& -                 						& -									 	\\
			FedNorm+															&																	& $0.680\,(0.784) \pm 0.285$  				& -5.3 									&																					& $\boldsymbol{0.942\,(0.946) \pm 0.023}$	& \textbf{6.4}$^\ast$				 	\\
			\cdashline{1-1}														\cdashline{3-4}\cdashline{6-7}
			Centralized															&																	& $0.744\,(0.893) \pm 0.317$  				& 3.6 									&																					& $0.938\,(0.950) \pm 0.035$				& 6.0								  	\\
 			\bottomrule
		\end{tabular}
	\end{adjustbox}
\end{table*}

Overall, it can be observed that Dice per patient scores up to $0.961$ for CT and up to $0.941$ for MRI are
reached on the client test sets. Hence, these results demonstrate that our proposed method for multi-modal
liver segmentation can be trained successfully on CT and MRI data with FL, by achieving a high performance.

\begin{figure*}[ht]
	\centering
		\resizebox{\textwidth}{!}{
			\includegraphics{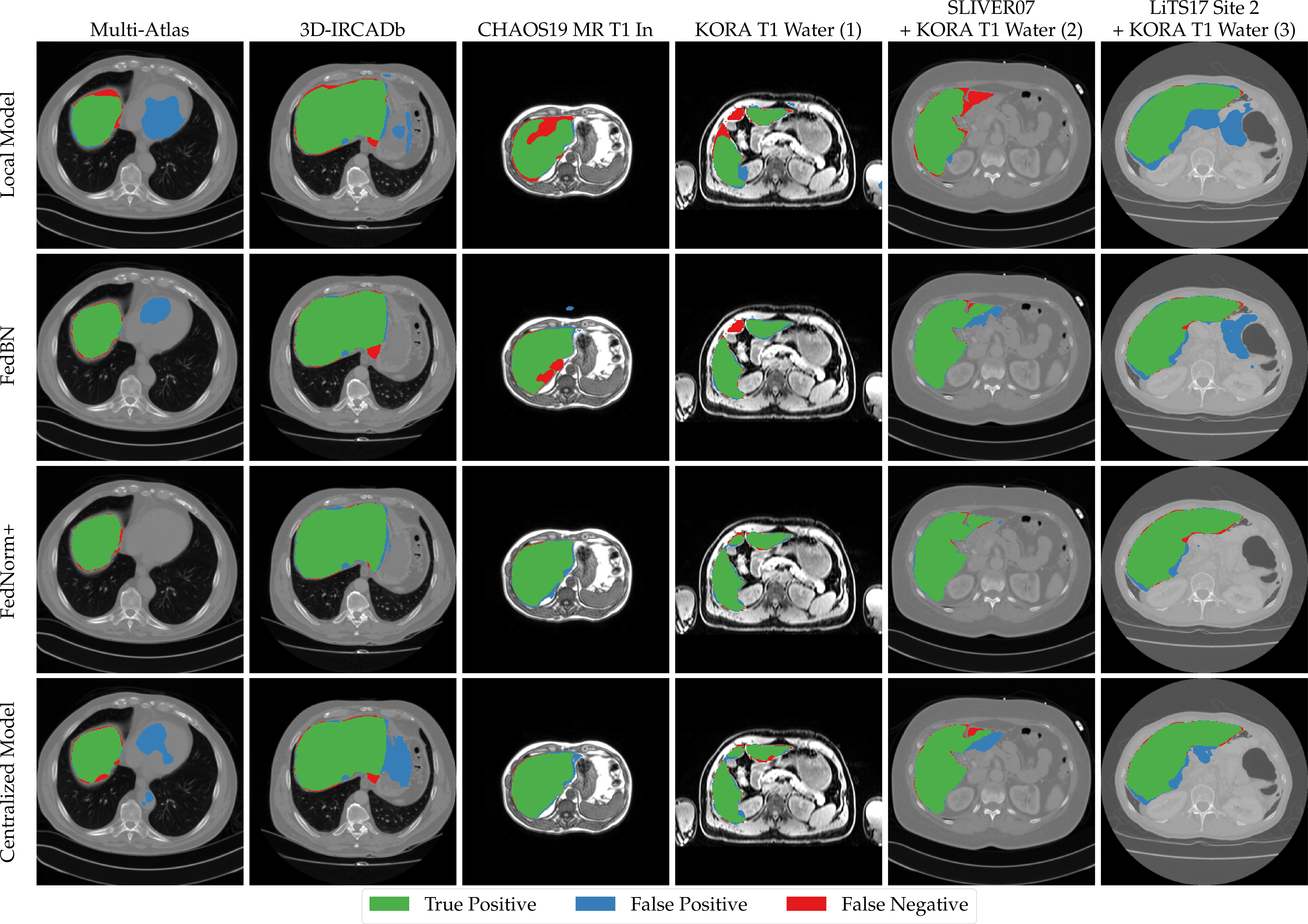}
		}
	\caption[Visual Results: Non-IID 2]{Comparison of visual results of the liver segmentation predictions from the local models, \fedbn, \fednormp, and the centralized model (rows) on one slice from the test set of each client (columns) of the Non-IID 2 setting.}\label{fig:visual-results}
\end{figure*}

To confirm these observations, a visual comparison of the performances of the local models, \fedbn, 
\fednormp, and the centralized model on the test sets of the clients in the Non-IID 2 setting is presented 
in \Cref{fig:visual-results}, as \fedbn can be identified as one of the strongest competitors to 
\fednormp according to \Cref{tab:comparison}. It can be seen that the model of \fednormp shows a good 
performance, where the local models fail and even where \fedbn and the centralized model do not provide 
precise liver segmentations. It also shows that \fednormp is able to accurately segment the liver even in 
unfamiliar appearances (KORA T1 Water (1) client) and that it does not falsely segment areas with similar 
intensity values as the liver, like the other models on the Multi-Atlas client.
\section{Conclusion}\label{sec:conclusion}
Based on our experimental results, we conclude that our proposed FL methods for multi-modal liver 
segmentation can be trained on data from multiple modalities (e.g. CT and MRI) by reaching a
high performance.
We have observed that the federated models in our experiments consistently
outperform local models, which are trained on each dataset individually.
Furthermore, in some cases, the federated models have also outperformed the 
centralized models. This demonstrates that it is not necessary to collect 
data from multiple hospitals and modalities at a central location (data 
lake) to train deep learning models for multi-modal liver segmentation on CT
and MRI data. It is rather sufficient to have a few participating clients 
(hospitals) inside the federation during FL, where each client contains data
from either only CT/MRI, or both modalities.
We have further observed that our approaches outperform other 
state-of-the-art FL algorithms for most of the clients in our FL settings. 
However, the problem of \fednorm is that it cannot handle
clients with data from both modalities. To this end, we have proposed
\fednormp, which normalizes the data based on 
slice-wise modality information and a hard-coded mode assignment.
According to our results, it is also sufficient to use a relatively small 
version of the U-Net~\citep{UNet} 
to achieve high performances for multi-modal liver segmentation.

\appendix
\bibliography{refs}

\begin{thebibliography}{51}
\providecommand{\natexlab}[1]{#1}
\providecommand{\url}[1]{\texttt{#1}}
\expandafter\ifx\csname urlstyle\endcsname\relax
  \providecommand{\doi}[1]{doi: #1}\else
  \providecommand{\doi}{doi: \begingroup \urlstyle{rm}\Url}\fi

\bibitem[3DI()]{3DIRCADb}
{3D-IRCADb}.
\newblock \url{https://www.ircad.fr/research/3dircadb/}.
\newblock (Accessed: 2021-01-07).

\bibitem[KOR()]{KORA}
{Kooperative Gesundheitsforschung in der Region Augsburg (KORA)}.
\newblock \url{https://www.helmholtz-muenchen.de/en/kora/index.html}.
\newblock (Accessed: 2021-05-21).

\bibitem[Andreux et~al.(2020)Andreux, du~Terrail, Beguier, and Tramel]{SiloBN}
Mathieu Andreux, Jean~Ogier du~Terrail, Constance Beguier, and Eric~W. Tramel.
\newblock Siloed federated learning for multi-centric histopathology datasets.
\newblock In Shadi Albarqouni, Spyridon Bakas, Konstantinos Kamnitsas, M.~Jorge
  Cardoso, Bennett Landman, Wenqi Li, Fausto Milletari, Nicola Rieke, Holger
  Roth, Daguang Xu, and Ziyue Xu, editors, \emph{Domain Adaptation and
  Representation Transfer, and Distributed and Collaborative Learning}, pages
  129--139, Cham, 2020. Springer International Publishing.
\newblock ISBN 978-3-030-60548-3.
\newblock \doi{https://doi.org/10.1007/978-3-030-60548-3_13}.

\bibitem[Asrani et~al.(2019)Asrani, Devarbhavi, Eaton, and
  Kamath]{LiverDiseases}
Sumeet~K. Asrani, Harshad Devarbhavi, John Eaton, and Patrick~S. Kamath.
\newblock Burden of liver diseases in the world.
\newblock \emph{Journal of Hepatology}, 70\penalty0 (1):\penalty0 151--171,
  2019.
\newblock ISSN 0168-8278.
\newblock \doi{https://doi.org/10.1016/j.jhep.2018.09.014}.

\bibitem[Bakhtiari et~al.(2020)Bakhtiari, Nasirigerdeh, Torkzadehmahani, Bayat,
  Blumenthal, List, and Baumbach]{FederatedMMB}
Mohammad Bakhtiari, Reza Nasirigerdeh, Reihaneh Torkzadehmahani, Amirhossein
  Bayat, David~B. Blumenthal, Markus List, and Jan Baumbach.
\newblock Federated multi-mini-batch: An efficient training approach to
  federated learning in non-iid environments, 2020.

\bibitem[Bilic et~al.(2019)Bilic, Christ, Vorontsov, Chlebus, Chen, Dou, Fu,
  Han, Heng, Hesser, Kadoury, Konopczynski, Le, Li, Li, Lipkovà, Lowengrub,
  Meine, Moltz, Pal, Piraud, Qi, Qi, Rempfler, Roth, Schenk, Sekuboyina,
  Vorontsov, Zhou, Hülsemeyer, Beetz, Ettlinger, Gruen, Kaissis, Lohöfer,
  Braren, Holch, Hofmann, Sommer, Heinemann, Jacobs, Mamani, van Ginneken,
  Chartrand, Tang, Drozdzal, Ben-Cohen, Klang, Amitai, Konen, Greenspan,
  Moreau, Hostettler, Soler, Vivanti, Szeskin, Lev-Cohain, Sosna, Joskowicz,
  and Menze]{LiTS17}
Patrick Bilic, Patrick~Ferdinand Christ, Eugene Vorontsov, Grzegorz Chlebus,
  Hao Chen, Qi~Dou, Chi-Wing Fu, Xiao Han, Pheng-Ann Heng, Jürgen Hesser,
  Samuel Kadoury, Tomasz Konopczynski, Miao Le, Chunming Li, Xiaomeng Li, Jana
  Lipkovà, John Lowengrub, Hans Meine, Jan~Hendrik Moltz, Chris Pal, Marie
  Piraud, Xiaojuan Qi, Jin Qi, Markus Rempfler, Karsten Roth, Andrea Schenk,
  Anjany Sekuboyina, Eugene Vorontsov, Ping Zhou, Christian Hülsemeyer, Marcel
  Beetz, Florian Ettlinger, Felix Gruen, Georgios Kaissis, Fabian Lohöfer,
  Rickmer Braren, Julian Holch, Felix Hofmann, Wieland Sommer, Volker
  Heinemann, Colin Jacobs, Gabriel Efrain~Humpire Mamani, Bram van Ginneken,
  Gabriel Chartrand, An~Tang, Michal Drozdzal, Avi Ben-Cohen, Eyal Klang,
  Marianne~M. Amitai, Eli Konen, Hayit Greenspan, Johan Moreau, Alexandre
  Hostettler, Luc Soler, Refael Vivanti, Adi Szeskin, Naama Lev-Cohain, Jacob
  Sosna, Leo Joskowicz, and Bjoern~H. Menze.
\newblock The liver tumor segmentation benchmark (lits), 2019.

\bibitem[Bishop(2006)]{Bishop}
Christopher~M. Bishop.
\newblock \emph{Pattern Recognition and Machine Learning (Information Science
  and Statistics)}.
\newblock Springer-Verlag, Berlin, Heidelberg, 2006.
\newblock ISBN 0387310738.

\bibitem[Chen et~al.(2020)Chen, Dou, Chen, Qin, and
  Heng]{MultiModalMedicalImageSegmentation}
Cheng Chen, Qi~Dou, Hao Chen, Jing Qin, and Pheng~Ann Heng.
\newblock Unsupervised bidirectional cross-modality adaptation via deeply
  synergistic image and feature alignment for medical image segmentation.
\newblock \emph{IEEE Transactions on Medical Imaging}, 39\penalty0
  (7):\penalty0 2494--2505, 2020.
\newblock \doi{10.1109/TMI.2020.2972701}.

\bibitem[Chen et~al.(2012)Chen, Wang, Hu, Zhao, and
  Wu]{LiverSegmentationGraphCut2}
Yufei Chen, Zhicheng Wang, Jinyong Hu, Weidong Zhao, and Qidi Wu.
\newblock The domain knowledge based graph-cut model for liver ct segmentation.
\newblock \emph{Biomedical Signal Processing and Control}, 7\penalty0
  (6):\penalty0 591--598, 2012.
\newblock ISSN 1746-8094.
\newblock \doi{https://doi.org/10.1016/j.bspc.2012.04.005}.
\newblock Biomedical Image Restoration and Enhancement.

\bibitem[Christ et~al.(2016)Christ, Elshaer, Ettlinger, Tatavarty, Bickel,
  Bilic, Rempfler, Armbruster, Hofmann, D'Anastasi, Sommer, Ahmadi, and
  Menze]{CascadedFCN}
Patrick~Ferdinand Christ, Mohamed Ezzeldin~A. Elshaer, Florian Ettlinger, Sunil
  Tatavarty, Marc Bickel, Patrick Bilic, Markus Rempfler, Marco Armbruster,
  Felix Hofmann, Melvin D'Anastasi, Wieland~H. Sommer, Seyed-Ahmad Ahmadi, and
  Bjoern~H. Menze.
\newblock Automatic liver and lesion segmentation in ct using cascaded fully
  convolutional neural networks and 3d conditional random fields.
\newblock In Sebastien Ourselin, Leo Joskowicz, Mert~R. Sabuncu, Gozde Unal,
  and William Wells, editors, \emph{Medical Image Computing and
  Computer-Assisted Intervention -- MICCAI 2016}, pages 415--423, Cham, 2016.
  Springer International Publishing.
\newblock ISBN 978-3-319-46723-8.
\newblock \doi{https://doi.org/10.1007/978-3-319-46723-8_48}.

\bibitem[Christ et~al.(2017)Christ, Ettlinger, Grün, Elshaera, Lipkova,
  Schlecht, Ahmaddy, Tatavarty, Bickel, Bilic, Rempfler, Hofmann, Anastasi,
  Ahmadi, Kaissis, Holch, Sommer, Braren, Heinemann, and
  Menze]{MMLTSCascadedFCN}
Patrick~Ferdinand Christ, Florian Ettlinger, Felix Grün, Mohamed Ezzeldin~A.
  Elshaera, Jana Lipkova, Sebastian Schlecht, Freba Ahmaddy, Sunil Tatavarty,
  Marc Bickel, Patrick Bilic, Markus Rempfler, Felix Hofmann, Melvin~D
  Anastasi, Seyed-Ahmad Ahmadi, Georgios Kaissis, Julian Holch, Wieland Sommer,
  Rickmer Braren, Volker Heinemann, and Bjoern Menze.
\newblock Automatic liver and tumor segmentation of ct and mri volumes using
  cascaded fully convolutional neural networks, 2017.

\bibitem[Couteaux et~al.(2021)Couteaux, Trintignac, Nempont, Pizaine,
  Vlachomitrou, Valette, Milot, and Bloch]{MMLSMRI}
Vincent Couteaux, Mathilde Trintignac, Olivier Nempont, Guillaume Pizaine,
  Anna~Sesilia Vlachomitrou, Pierre-Jean Valette, Laurent Milot, and Isabelle
  Bloch.
\newblock Comparing deep learning strategies for paired but unregistered
  multimodal segmentation of the liver in t1 and t2-weighted mri, 2021.

\bibitem[Deecke et~al.(2018)Deecke, Murray, and Bilen]{ModeNormalization}
Lucas Deecke, Iain Murray, and Hakan Bilen.
\newblock Mode normalization, 2018.
\newblock {International Conference on Learning Representations (2019)}.

\bibitem[Dice(1945)]{Dice}
Lee~R. Dice.
\newblock Measures of the amount of ecologic association between species.
\newblock \emph{Ecology}, 26\penalty0 (3):\penalty0 297--302, 1945.
\newblock \doi{https://doi.org/10.2307/1932409}.
\newblock URL
  \url{https://esajournals.onlinelibrary.wiley.com/doi/abs/10.2307/1932409}.

\bibitem[Dixon(1984)]{DixonMethod}
WT~Dixon.
\newblock Simple proton spectroscopic imaging.
\newblock \emph{Radiology}, 153\penalty0 (1):\penalty0 189—194, October 1984.
\newblock ISSN 0033-8419.
\newblock \doi{10.1148/radiology.153.1.6089263}.
\newblock URL \url{https://doi.org/10.1148/radiology.153.1.6089263}.

\bibitem[Global Cancer Observatory()]{LiverCancer}
Global Cancer Observatory.
\newblock {World Fact Sheets}.
\newblock
  \url{https://gco.iarc.fr/today/data/factsheets/populations/900-world-fact-sheets.pdf},
  2021.
\newblock (Accessed: 2021-05-29).

\bibitem[Heimann et~al.(2009)Heimann, van Ginneken, Styner, Arzhaeva, Aurich,
  Bauer, Beck, Becker, Beichel, Bekes, Bello, Binnig, Bischof, Bornik, Cashman,
  Chi, Cordova, Dawant, Fidrich, Furst, Furukawa, Grenacher, Hornegger,
  KainmÜller, Kitney, Kobatake, Lamecker, Lange, Lee, Lennon, Li, Li, Meinzer,
  Nemeth, Raicu, Rau, van Rikxoort, Rousson, Rusko, Saddi, Schmidt, Seghers,
  Shimizu, Slagmolen, Sorantin, Soza, Susomboon, Waite, Wimmer, and
  Wolf]{SLIVER07}
Tobias Heimann, Bram van Ginneken, Martin~A. Styner, Yulia Arzhaeva, Volker
  Aurich, Christian Bauer, Andreas Beck, Christoph Becker, Reinhard Beichel,
  GyÖrgy Bekes, Fernando Bello, Gerd Binnig, Horst Bischof, Alexander Bornik,
  Peter M.~M. Cashman, Ying Chi, AndrÉs Cordova, Benoit~M. Dawant, MÁrta
  Fidrich, Jacob~D. Furst, Daisuke Furukawa, Lars Grenacher, Joachim Hornegger,
  Dagmar KainmÜller, Richard~I. Kitney, Hidefumi Kobatake, Hans Lamecker,
  Thomas Lange, Jeongjin Lee, Brian Lennon, Rui Li, Senhu Li, Hans-Peter
  Meinzer, GÁbor Nemeth, Daniela~S. Raicu, Anne-Mareike Rau, Eva~M. van
  Rikxoort, MikaËl Rousson, LÁszlÓ Rusko, Kinda~A. Saddi, GÜnter Schmidt,
  Dieter Seghers, Akinobu Shimizu, Pieter Slagmolen, Erich Sorantin, Grzegorz
  Soza, Ruchaneewan Susomboon, Jonathan~M. Waite, Andreas Wimmer, and Ivo Wolf.
\newblock Comparison and evaluation of methods for liver segmentation from ct
  datasets.
\newblock \emph{IEEE Transactions on Medical Imaging}, 28\penalty0
  (8):\penalty0 1251--1265, 2009.
\newblock \doi{10.1109/TMI.2009.2013851}.

\bibitem[Hsu et~al.(2019)Hsu, Qi, and Brown]{FedAvgM}
Tzu-Ming~Harry Hsu, Hang Qi, and Matthew Brown.
\newblock Measuring the effects of non-identical data distribution for
  federated visual classification, 2019.

\bibitem[Hsu et~al.(2020)Hsu, Qi, and Brown]{FedVC}
Tzu-Ming~Harry Hsu, Hang Qi, and Matthew Brown.
\newblock Federated visual classification with real-world data distribution.
\newblock In Andrea Vedaldi, Horst Bischof, Thomas Brox, and Jan-Michael Frahm,
  editors, \emph{Computer Vision -- ECCV 2020}, pages 76--92, Cham, 2020.
  Springer International Publishing.
\newblock ISBN 978-3-030-58607-2.
\newblock \doi{https://doi.org/10.1007/978-3-030-58607-2_5}.

\bibitem[Ioffe and Szegedy(2015)]{BatchNormalization}
Sergey Ioffe and Christian Szegedy.
\newblock Batch normalization: Accelerating deep network training by reducing
  internal covariate shift.
\newblock In Francis Bach and David Blei, editors, \emph{Proceedings of the
  32nd International Conference on Machine Learning}, volume~37 of
  \emph{Proceedings of Machine Learning Research}, pages 448--456, Lille,
  France, Jul 2015. PMLR.

\bibitem[Isensee et~al.(2018)Isensee, Petersen, Klein, Zimmerer, Jaeger, Kohl,
  Wasserthal, Koehler, Norajitra, Wirkert, and Maier-Hein]{nnUNet}
Fabian Isensee, Jens Petersen, Andre Klein, David Zimmerer, Paul~F. Jaeger,
  Simon Kohl, Jakob Wasserthal, Gregor Koehler, Tobias Norajitra, Sebastian
  Wirkert, and Klaus~H. Maier-Hein.
\newblock nnu-net: Self-adapting framework for u-net-based medical image
  segmentation, 2018.

\bibitem[Jiang and
  Veeraraghavan(2020)]{CrossModalityUnsupervisedOrganSegmentation}
Jue Jiang and Harini Veeraraghavan.
\newblock Unified cross-modality feature disentangler for unsupervised
  multi-domain mri abdomen organs segmentation.
\newblock In Anne~L. Martel, Purang Abolmaesumi, Danail Stoyanov, Diana Mateus,
  Maria~A. Zuluaga, S.~Kevin Zhou, Daniel Racoceanu, and Leo Joskowicz,
  editors, \emph{Medical Image Computing and Computer Assisted Intervention --
  MICCAI 2020}, pages 347--358, Cham, 2020. Springer International Publishing.
\newblock ISBN 978-3-030-59713-9.
\newblock \doi{https://doi.org/10.1007/978-3-030-59713-9_34}.

\bibitem[Kaissis et~al.(2021)Kaissis, Ziller, Passerat-Palmbach, Ryffel,
  Usynin, Trask, Lima, Mancuso, Jungmann, Steinborn, Saleh, Makowski, Rueckert,
  and Braren]{PrivacyPreservingFLMI}
Georgios Kaissis, Alexander Ziller, Jonathan Passerat-Palmbach, Th{\'e}o
  Ryffel, Dmitrii Usynin, Andrew Trask, Ion{\'e}sio Lima, Jason Mancuso,
  Friederike Jungmann, Marc-Matthias Steinborn, Andreas Saleh, Marcus Makowski,
  Daniel Rueckert, and Rickmer Braren.
\newblock End-to-end privacy preserving deep learning on multi-institutional
  medical imaging.
\newblock \emph{Nature Machine Intelligence}, 3\penalty0 (6):\penalty0
  473--484, Jun 2021.
\newblock ISSN 2522-5839.
\newblock \doi{10.1038/s42256-021-00337-8}.
\newblock URL \url{https://doi.org/10.1038/s42256-021-00337-8}.

\bibitem[Kaissis et~al.(2020)Kaissis, Makowski, R{\"u}ckert, and
  Braren]{PrivacyPreservingFLMI2}
Georgios~A. Kaissis, Marcus~R. Makowski, Daniel R{\"u}ckert, and Rickmer~F.
  Braren.
\newblock Secure, privacy-preserving and federated machine learning in medical
  imaging.
\newblock \emph{Nature Machine Intelligence}, 2\penalty0 (6):\penalty0
  305--311, Jun 2020.
\newblock ISSN 2522-5839.
\newblock \doi{10.1038/s42256-020-0186-1}.
\newblock URL \url{https://doi.org/10.1038/s42256-020-0186-1}.

\bibitem[Kavur et~al.(2019)Kavur, Selver, Dicle, Barış, and Gezer]{CHAOS19}
Ali~Emre Kavur, M.~Alper Selver, Oğuz Dicle, Mustafa Barış, and N.~Sinem
  Gezer.
\newblock {CHAOS - Combined (CT-MR) Healthy Abdominal Organ Segmentation
  Challenge Data}, Apr 2019.
\newblock URL \url{https://doi.org/10.5281/zenodo.3362844}.

\bibitem[Kingma and Ba(2015)]{Adam}
Diederik~P. Kingma and Jimmy Ba.
\newblock Adam: {A} method for stochastic optimization.
\newblock In Yoshua Bengio and Yann LeCun, editors, \emph{3rd International
  Conference on Learning Representations, {ICLR} 2015, San Diego, CA, USA, May
  7-9, 2015, Conference Track Proceedings}, 2015.

\bibitem[Kodym et~al.(2019)Kodym, {\v{S}}pan{\v{e}}l, and Herout]{DiceLoss}
Old{\v{r}}ich Kodym, Michal {\v{S}}pan{\v{e}}l, and Adam Herout.
\newblock Segmentation of head and neck organs at risk using cnn with batch
  dice loss.
\newblock In Thomas Brox, Andr{\'e}s Bruhn, and Mario Fritz, editors,
  \emph{Pattern Recognition}, pages 105--114, Cham, 2019. Springer
  International Publishing.
\newblock ISBN 978-3-030-12939-2.

\bibitem[Kravchenko and Bullock(1999)]{RelativeImprovement}
Alexandra Kravchenko and Donald~G. Bullock.
\newblock A comparative study of interpolation methods for mapping soil
  properties.
\newblock \emph{Agronomy Journal}, 91\penalty0 (3):\penalty0 393--400, 1999.
\newblock \doi{https://doi.org/10.2134/agronj1999.00021962009100030007x}.

\bibitem[Landman et~al.(2013)Landman, Xu, Igelsias, Styner, Langerak, and
  Klein]{MultiAtlas}
Bennett Landman, Zhoubing Xu, Juan~Eugenio Igelsias, Martin Styner,
  Thomas~Robin Langerak, and Arno Klein.
\newblock {Multi-Atlas Labeling Beyond the Cranial Vault - Workshop and
  Challenge}.
\newblock \url{https://www.synapse.org/\#!Synapse:syn3193805/wiki/89480}, 2013.
\newblock (Accessed: 2021-01-03).

\bibitem[Li et~al.(2020{\natexlab{a}})Li, Sahu, Zaheer, Sanjabi, Talwalkar, and
  Smith]{FedProx}
Tian Li, Anit~Kumar Sahu, Manzil Zaheer, Maziar Sanjabi, Ameet Talwalkar, and
  Virginia Smith.
\newblock Federated optimization in heterogeneous networks.
\newblock In I.~Dhillon, D.~Papailiopoulos, and V.~Sze, editors,
  \emph{Proceedings of Machine Learning and Systems}, volume~2, pages 429--450,
  2020{\natexlab{a}}.

\bibitem[Li et~al.(2019)Li, Milletar{\`i}, Xu, Rieke, Hancox, Zhu, Baust,
  Cheng, Ourselin, Cardoso, and Feng]{PrivacyPreservingFL}
Wenqi Li, Fausto Milletar{\`i}, Daguang Xu, Nicola Rieke, Jonny Hancox, Wentao
  Zhu, Maximilian Baust, Yan Cheng, S{\'e}bastien Ourselin, M.~Jorge Cardoso,
  and Andrew Feng.
\newblock Privacy-preserving federated brain tumour segmentation.
\newblock In Heung-Il Suk, Mingxia Liu, Pingkun Yan, and Chunfeng Lian,
  editors, \emph{Machine Learning in Medical Imaging}, pages 133--141, Cham,
  2019. Springer International Publishing.
\newblock ISBN 978-3-030-32692-0.
\newblock \doi{https://doi.org/10.1007/978-3-030-32692-0_16}.

\bibitem[Li et~al.(2020{\natexlab{b}})Li, Huang, Yang, Wang, and
  Zhang]{FedAvgNonIID}
Xiang Li, Kaixuan Huang, Wenhao Yang, Shusen Wang, and Zhihua Zhang.
\newblock On the convergence of fedavg on non-iid data.
\newblock In \emph{International Conference on Learning Representations},
  2020{\natexlab{b}}.
\newblock URL \url{https://openreview.net/forum?id=HJxNAnVtDS}.

\bibitem[Li et~al.(2018)Li, Chen, Qi, Dou, Fu, and Heng]{HDenseUNet}
Xiaomeng Li, Hao Chen, Xiaojuan Qi, Qi~Dou, Chi-Wing Fu, and Pheng-Ann Heng.
\newblock H-denseunet: Hybrid densely connected unet for liver and tumor
  segmentation from ct volumes.
\newblock \emph{IEEE Transactions on Medical Imaging}, 37\penalty0
  (12):\penalty0 2663--2674, 2018.
\newblock \doi{10.1109/TMI.2018.2845918}.

\bibitem[Li et~al.(2020{\natexlab{c}})Li, Gu, Dvornek, Staib, Ventola, and
  Duncan]{PrivacyPreservingFL2}
Xiaoxiao Li, Yufeng Gu, Nicha Dvornek, Lawrence~H. Staib, Pamela Ventola, and
  James~S. Duncan.
\newblock Multi-site fmri analysis using privacy-preserving federated learning
  and domain adaptation: Abide results.
\newblock \emph{Medical Image Analysis}, 65:\penalty0 101765,
  2020{\natexlab{c}}.
\newblock ISSN 1361-8415.
\newblock \doi{https://doi.org/10.1016/j.media.2020.101765}.

\bibitem[Li et~al.(2021)Li, JIANG, Zhang, Kamp, and Dou]{FedBN}
Xiaoxiao Li, Meirui JIANG, Xiaofei Zhang, Michael Kamp, and Qi~Dou.
\newblock Fed{BN}: Federated learning on non-{IID} features via local batch
  normalization.
\newblock In \emph{International Conference on Learning Representations}.
  OpenReview.net, 2021.
\newblock URL \url{https://openreview.net/forum?id=6YEQUn0QICG}.

\bibitem[Lu et~al.(2017)Lu, Wu, Hu, Peng, and Kong]{LiverSegmentationGraphCut}
Fang Lu, Fa~Wu, Peijun Hu, Zhiyi Peng, and Dexing Kong.
\newblock Automatic 3d liver location and segmentation via convolutional neural
  networks and graph cut.
\newblock \emph{International Journal of Computer Assisted Radiology and
  Surgery}, 12\penalty0 (2):\penalty0 171--182, Feb 2017.
\newblock ISSN 1861-6429.
\newblock \doi{10.1007/s11548-016-1467-3}.

\bibitem[Ma(2008)]{DixonTechniques}
Jingfei Ma.
\newblock Dixon techniques for water and fat imaging.
\newblock \emph{Journal of Magnetic Resonance Imaging}, 28:\penalty0 543 --
  558, 09 2008.
\newblock \doi{10.1002/jmri.21492}.

\bibitem[McMahan et~al.(2017)McMahan, Moore, Ramage, Hampson, and
  Arcas]{FedAvg}
Brendan McMahan, Eider Moore, Daniel Ramage, Seth Hampson, and Blaise Aguera~y
  Arcas.
\newblock {Communication\hyp{}Efficient Learning of Deep Networks from
  Decentralized Data}.
\newblock In Aarti Singh and Jerry Zhu, editors, \emph{Proceedings of the 20th
  International Conference on Artificial Intelligence and Statistics},
  volume~54 of \emph{Proceedings of Machine Learning Research}, pages
  1273--1282. PMLR, Apr 2017.

\bibitem[Milletari et~al.(2016)Milletari, Navab, and Ahmadi]{VNet}
Fausto Milletari, Nassir Navab, and Seyed-Ahmad Ahmadi.
\newblock V-net: Fully convolutional neural networks for volumetric medical
  image segmentation.
\newblock In \emph{2016 Fourth International Conference on 3D Vision (3DV)},
  pages 565--571, 2016.
\newblock \doi{10.1109/3DV.2016.79}.

\bibitem[Mulay et~al.(2020)Mulay, Deepika, Jeevakala, Ram, and
  Sivaprakasam]{MMLS}
Supriti Mulay, G.~Deepika, S.~Jeevakala, Keerthi Ram, and Mohanasankar
  Sivaprakasam.
\newblock Liver segmentation from multimodal images using hed-mask r-cnn.
\newblock In Quanzheng Li, Richard Leahy, Bin Dong, and Xiang Li, editors,
  \emph{Multiscale Multimodal Medical Imaging}, pages 68--75, Cham, 2020.
  Springer International Publishing.
\newblock ISBN 978-3-030-37969-8.
\newblock \doi{https://doi.org/10.1007/978-3-030-37969-8_9}.

\bibitem[Oliva and Saini(2004)]{LiverCancerImagingModalities}
Maria~Raquel Oliva and Sanjay Saini.
\newblock Liver cancer imaging: role of ct, mri, us and pet.
\newblock \emph{Cancer imaging}, 4 Spec No A:\penalty0 42--46, Apr 2004.
\newblock ISSN 1470-7330.
\newblock \doi{10.1102/1470-7330.2004.0011}.
\newblock 18215974[pmid].

\bibitem[Ronneberger et~al.(2015)Ronneberger, Fischer, and Brox]{UNet}
Olaf Ronneberger, Philipp Fischer, and Thomas Brox.
\newblock U-net: Convolutional networks for biomedical image segmentation.
\newblock In Nassir Navab, Joachim Hornegger, William~M. Wells, and
  Alejandro~F. Frangi, editors, \emph{Medical Image Computing and
  Computer-Assisted Intervention -- MICCAI 2015}, pages 234--241, Cham, 2015.
  Springer International Publishing.
\newblock ISBN 978-3-319-24574-4.

\bibitem[Sheller et~al.(2020)Sheller, Edwards, Reina, Martin, Pati, Kotrotsou,
  Milchenko, Xu, Marcus, Colen, and Bakas]{MedicalFL}
Micah Sheller, Brandon Edwards, G.~Reina, Jason Martin, Sarthak Pati,
  Aikaterini Kotrotsou, Mikhail Milchenko, Weilin Xu, Daniel Marcus, Rivka
  Colen, and Spyridon Bakas.
\newblock Federated learning in medicine: facilitating multi-institutional
  collaborations without sharing patient data.
\newblock \emph{Scientific Reports}, 10, 07 2020.
\newblock \doi{10.1038/s41598-020-69250-1}.

\bibitem[Stan and Rostami(2021)]{Preprocessing}
Serban Stan and Mohammad Rostami.
\newblock Privacy preserving domain adaptation for semantic segmentation of
  medical images, 2021.

\bibitem[Sun et~al.(2017)Sun, Shrivastava, Singh, and Gupta]{TrainingDataSize}
Chen Sun, Abhinav Shrivastava, Saurabh Singh, and Abhinav Gupta.
\newblock Revisiting unreasonable effectiveness of data in deep learning era.
\newblock In \emph{2017 IEEE International Conference on Computer Vision
  (ICCV)}, pages 843--852, 2017.
\newblock \doi{10.1109/ICCV.2017.97}.

\bibitem[Valindria et~al.(2018)Valindria, Pawlowski, Rajchl, Lavdas, Aboagye,
  Rockall, Rueckert, and Glocker]{MMOS}
Vanya~V. Valindria, Nick Pawlowski, Martin Rajchl, Ioannis Lavdas, Eric~O.
  Aboagye, Andrea~G. Rockall, Daniel Rueckert, and Ben Glocker.
\newblock Multi-modal learning from unpaired images: Application to multi-organ
  segmentation in ct and mri.
\newblock In \emph{2018 IEEE Winter Conference on Applications of Computer
  Vision (WACV)}, pages 547--556, 2018.
\newblock \doi{10.1109/WACV.2018.00066}.
\newblock © 2011 IEEE.

\bibitem[Wang et~al.(2019)Wang, Mamidipalli, Retson, Bahrami, Hasenstab,
  Blansit, Bass, Delgado, Cunha, Middleton, Loomba, Neuschwander-Tetri, Sirlin,
  and Hsiao]{MMLSGCNN}
Kang Wang, Adrija Mamidipalli, Tara Retson, Naeim Bahrami, Kyle Hasenstab,
  Kevin Blansit, Emily Bass, Timoteo Delgado, Guilherme Cunha, Michael
  Middleton, Rohit Loomba, Brent Neuschwander-Tetri, Claude Sirlin, and Albert
  Hsiao.
\newblock Automated ct and mri liver segmentation and biometry using a
  generalized convolutional neural network.
\newblock \emph{Radiology: Artificial Intelligence}, 1:\penalty0 180022, 03
  2019.
\newblock \doi{10.1148/ryai.2019180022}.

\bibitem[Wang et~al.(2020)Wang, Kumar, and Chandra]{DataHeterogeneity}
Yuanli Wang, Dhruv Kumar, and Abhishek Chandra.
\newblock Poster: Exploiting data heterogeneity for performance and reliability
  in federated learning.
\newblock In \emph{2020 IEEE/ACM Symposium on Edge Computing (SEC)}, pages
  164--166, 2020.
\newblock \doi{10.1109/SEC50012.2020.00023}.

\bibitem[Xue et~al.(2021)Xue, Li, Zhang, Lu, Zhu, Shen, Shah, and
  Bennamoun]{MultiModalLiverLesionSegmentation}
Zhongliang Xue, Ping Li, Liang Zhang, Xiaoyuan Lu, Guangming Zhu, Peiyi Shen,
  Syed Afaq~Ali Shah, and Mohammed Bennamoun.
\newblock Multi-modal co-learning for liver lesion segmentation on pet-ct
  images.
\newblock \emph{IEEE Transactions on Medical Imaging}, pages 1--1, 2021.
\newblock \doi{10.1109/TMI.2021.3089702}.

\bibitem[Yang et~al.(2019)Yang, Dvornek, Zhang, Chapiro, Lin, and Duncan]{DADR}
Junlin Yang, Nicha~C. Dvornek, Fan Zhang, Julius Chapiro, MingDe Lin, and
  James~S. Duncan.
\newblock Unsupervised domain adaptation via disentangled representations:
  Application to cross-modality liver segmentation.
\newblock In Dinggang Shen, Tianming Liu, Terry~M. Peters, Lawrence~H. Staib,
  Caroline Essert, Sean Zhou, Pew-Thian Yap, and Ali Khan, editors,
  \emph{Medical Image Computing and Computer Assisted Intervention -- MICCAI
  2019}, pages 255--263, Cham, 2019. Springer International Publishing.
\newblock ISBN 978-3-030-32245-8.
\newblock \doi{https://doi.org/10.1007/978-3-030-32245-8_29}.

\bibitem[Zhao et~al.(2018)Zhao, Li, Lai, Suda, Civin, and Chandra]{FLNonIID}
Yue Zhao, Meng Li, Liangzhen Lai, Naveen Suda, Damon Civin, and Vikas Chandra.
\newblock Federated learning with non-iid data, 2018.

\end{thebibliography}

\end{document}